\documentclass[twoside, a4paper, leqno]{article}

\usepackage{amsmath, amssymb, amsthm}
\usepackage{mathtools}
\usepackage{mathrsfs}
\usepackage[sort&compress, square, comma, numbers]{natbib} 
\usepackage{hyperref}
\usepackage{bm}
\usepackage{graphicx}        
\usepackage{multicol}        
\usepackage[bottom]{footmisc}
\usepackage{comment}
\usepackage{geometry}

\graphicspath{{FIGURES/}
 {pics/}}

\begin{document}
\title{Reconstruction of annular bi-layered
  media in cylindrical waveguide section} 

\author{Anders Eriksson
 \thanks{FMC Technologies, Kongsberg Norway,
 e-mail: \texttt{danandersgustav@fmcti.com}}
\and Truls Martin Larsen
 \thanks{FMC Technologies, Kongsberg Norway,
 e-mail: \texttt{trulsmartin.larsen@fmcti.com}}
\and Larisa Beilina
 \thanks{Department of Mathematical Sciences, Chalmers University of Technology and  University of Gothenburg, SE-412 96 Gothenburg, Sweden,
 e-mail: \texttt{larisa@chalmers.se}}}

\date{}
\maketitle

\begin{abstract}
A radial transverse resonance model for two cylindrical concentric
layers with different complex dielectric constants is presented. An
inverse problem with four unknowns - 3 physical material parameters
and one dimensional dielectric layer thickness parameter- is solved by
employing TE110 and TE210 modes with different radial field
distribution. First a Newton-Raphson algorithm is used to solve a
least square problem with a Lorentzian function (as resonance model
and "measured" data generator). Then found resonance frequencies and
quality factors are used in a second inverse Newton-Raphson algorithm
that solves four transverse resonance equations in order to get four
unknown parameters. The use of TE110 and TE210  models offers one
dimensional radial tomographic capability.
An open ended coax quarter-wave resonator is added to the sensor
topology, and the effect on the convergence is investigated.

\end{abstract}

\section{\label{sec:intro}Introduction}

Extraction of material parameters and/or dimensions based on
distributed resonator measurements has been around for
decades. Characterization of distributed microwave resonators
dielectric material from resonance frequency and quality factor
measurements is found in \cite{Eriksson}. A comparison of inverse
methods for extracting resonant frequency and quality factor is given
in \cite{Petersan}. Typically, the dielectric filling of the
resonators is homogeneous, but there are not really any restrictions
for allowing inhomogeneous dielectric filling. In this work, two
annular concentric cylindrical layers which are enclosed in a finite
conductive metallic pipe, each with a frequency dependent dielectric
constant, is modelled and four unknown physical parameters are found
using a transverse resonance radial model, alone as well as with
additional open ended coax quarter wave resonators. The possible
applications are characterization of metallic pipes with annular flow,
or characterization of optical fibers and similar geometries.

\section{Notations}

 \label{sec:theory} 
The complex dielectric constant can be expressed as:
\begin{equation}\label{eq1}
\varepsilon =  \varepsilon_0 \varepsilon_r = \varepsilon_0 (\varepsilon_{Re } - i \varepsilon_{Im }) =
  \varepsilon_0 \left (\varepsilon_{Re } - i\frac{\sigma}{\omega} \right),
\end{equation}
where $\varepsilon_0$ is the electric permittivity in vacuum, 
$\varepsilon_r$ is the dimensionless relative electric permittivity,  
$\varepsilon_{Re }$, $\varepsilon_{Im }$ are the real and imaginary
parts of the relative permittivity for an arbitrary material,
respectively, $\sigma$ is the electrical conductivity, and $\omega$ is
angular frequency. The equation   (\ref{eq1}) applies to imperfect metal with finite
conductivity as well as to the dielectric materials under investigation.

\begin{table}[ht]
  \begin{tabular}{|p{0.3\linewidth}|p{0.6\linewidth}|}
    \hline
    \textbf{Measurement Parameter} & \textbf{Description} \\
    \hline
    h & Thickness of liquid layer - liquid having arbitrary mixture ratio
    (Water Void Fraction) of condensate and saline water. \\
    \hline
    $R_{WLR}$ & Water Liquid Ratio (WLR) - water fraction in liquid. \\ 
    \hline
    $R_{DGR}$ & Droplet Gas Ratio  (DGR) - ratio of liquid immersed in droplet form in
    gas continuous volume (note that WLR and Salinity is the same in
    liquid film as well as in droplets). \\
    \hline
    $s$ & Salinity - salt concentration in water. \\
    \hline
  \end{tabular}
  \label{tab:measumentparameters}
  \caption{Parameters  to be determined.}
\end{table}

\begin{table}[ht]
  \begin{tabular}{|p{0.3\linewidth}|p{0.6\linewidth}|}
    \hline
    \textbf{Parameter} & \textbf{Description} \\
    \hline
    $\varepsilon_r$ & Relative electric permittivity \\
    \hline
    $\varepsilon_0$ & Electric permittivity in vacuum \\
    \hline
    $\varepsilon$ & $\varepsilon_0 \cdot \varepsilon_r$ \\
    \hline
    $\mu_r$ & Relative magnetic permeability \\
    \hline
    $\mu_0$ & Magnetic permeability in vacuum \\
    \hline
    $\mu$ & $\mu_r \cdot \mu_0$\\
    \hline
    $\omega$ & Angular frequency \\
    \hline
    $\omega_0$ & Angular resonant frequency, $2\pi f_0$.\\
    \hline
    $f_0$ & Resonant frequency \\
    \hline
    $Q_0$ & Unloaded quality factor \\
    \hline
    $TE_{mnl}$ &  Transverse electric mode  with angular index m, radial
    index n, and longitudinal index z. \\
    \hline
    $TM$ & Transverse magnetic mode. \\
    \hline
    $H^{(1)}_{m}(r)$ & Hankel function of the first kind. \\
    \hline
    $H^{(2)}_{m}(r)$ & Hankel function of the second kind. \\
    \hline
    $H^{(1)\prime}_{m}(r)$ & Derivative of Hankel function of the first kind. \\
    \hline
    $H^{(2)\prime}_{m}(r)$ & Derivative of Hankel function of the second kind. \\
    \hline
    $J_m(r)$ & Bessel function. \\
    \hline
    $f_{Re} $ & Real part of the function f. \\
    \hline
    $f_{Im} $ & Imaginary part of the function f. \\
    \hline
    $S_{11}$ & Reflecton coefficient \\
    \hline
    $S_{21}$ & Transmission coefficient \\
    \hline
  \end{tabular}
  \label{tab:fundamentalparameters}
  \caption{Fundamental notations.}
\end{table}

\begin{figure}
\begin{center}
  \includegraphics[width= 0.5\linewidth]{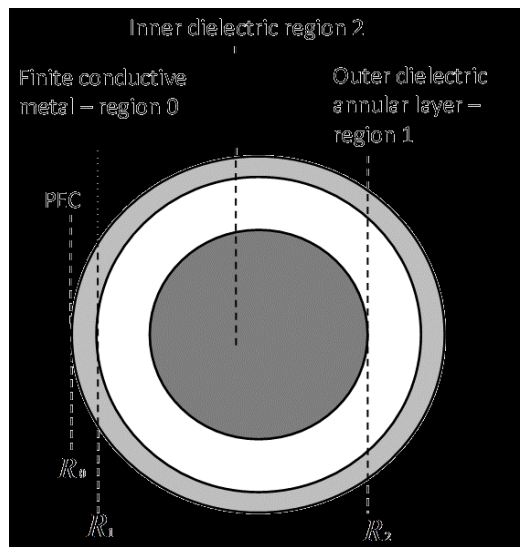}
\end{center}
  \caption{Cross-section of the annular waveguide section.}
\label{fig:cross_section}
\end{figure}

\subsection{\label{sec:emulsionsalinity} Modelling of permittivity of emulsions and saline water}
In this work, there are 4 different media (3 different media) - salt
(NaCl), hydrocarbon gas, Water ($H_2O$) and oil. No water vapor is
assumed in the calculations. 
Bruggemans model for emulsion permittivity \cite{mix_permitt_models,
  Bruggeman} is used. This model is applied for saline water mixed with
oil as well as for liquid droplets in gas. For the latter, Bruggemans
model is used twice - first for water mixed with oil in liquid
droplets, and then again to calculate the effective permittivity of
liquid droplets in gas. 

Several models  for effective complex permittivity of metal
powders in insulating dielectrics (e.g. teflon) are studied in 
 \cite{mix_permitt_models}, and one conclusion is
that Bruggeman models are relatively accurate for predicting real part
of permittivity while the well-known Maxwell-Garnett models have
higher accuracies for the imaginary part. This is relevant for the
case of an oil continuous regime with water content of high salinity -
where the electrically conducting saline water droplets are comparable
to electrically conducting metal powder.
Gadanis model \cite{Gadani} for saline water is used to describe a complex
permittivity of the saline water. The complex permittivity is a function
of salinity ($s$), temperature ($T$) and frequency ($f$) in this
model.

Thus, by applying these permittivity models with corresponding
fractions for each medium  (assuming that the known chemical
substances are  presented  but with unknown ratios),
the parameters are
reduced in this case to 4 unknowns, while in the case of using
directly a frequency-dependent complex permittivity for the liquid,
and a another frequency-dependent complex permittivity for the gas, 4
unknowns would be left to solve just for one frequency point along
with the liquid thickness, rendering 5 unknown. Then from these
permittivities, the WLR and droplet fraction in gas would have to be
found.  

For simplicity, the dielectric constants for gas and oil are set to 
be 1.7 and 2.0, respectively.

\subsection{\label{sec:direct}Direct Problem}
A full wave RF resonance model using transverse resonance method is
very computationally efficient and compact (equation wise). It
can model infinite long cylindrical pipe or waveguide, filled with
arbitrary concentric layers. Ideal boundary conditions in z-direction
may also be modeled.  

Full wave models for open ended coax are typically more complex and
require much more computational effort. In order to be used in a real
time application, a parameterized fast model without any numerical
integration would typically be needed.

\subsubsection{\label{sec:trans_res_method}Radial transverse resonance method}
For circular cylindrical waveguides composed of at least 2 different
dielectrics ($\varepsilon_{r1} \neq \varepsilon_{r2}$ and/or
$\mu_{r1} \neq \mu_{r1}$) supports in general no pure TM and TE modes 
except for symmetrical cases when angular index $m=0$ ($TE_{0nl}$ and
$TM_{0nl}$). In this case when angular index $m=0$, the characteristic
equation from the determinant of a 4x4 matrix (matrix derived from
matching tangential magnetic and electric fields $H_z, H_\theta, E_z$
and $E_\theta$ for the 2 layer dielectrics case) can be factorized 
such that the equation can be factorized to a product of a TM and TE
characteristic equation \cite{Harrington}. However, the elements in
the 4x4 matrix (for two layered circular cylindrical waveguide)
containing angular index $m$ also contains longitudinal wave number
$k_z$. At radial resonance (i.e. cut-off frequency), the longitudinal
wave number $k_z = 0$, and thus the same matrix elements that becomes
zero when $m=0$ also becomes zero when even for longitudinal index 
$k_z=0$. Thus, TM and TE radial resonances (where $k_z=0$) for any
angular index may be accommodated.

The quality factor of a TE or TM mode will in theory be decreased by
leaking waves, e.g. surface waves. Typically, for transmission based
resonator measurements, the external coupling to the resonator is
relatively weak - thus, $ |S_{11}|^2 + |S_{21}|^2 \approx 1$,
indicating that the unloaded quality factor (stored energy to
dissipated energy ratio) is dominant. In practice, 
this is realized by relatively small antennas/probes such that mainly
fringing antenna field is exciting the resonance. Under such
conditions, external quality factor is relatively large compared to
the unloaded quality factor, rendering unloaded $Q$ (intrinsic desired
resonator $Q$) equal to the loaded $Q$ (measured $Q$).  

The radial propagation constant {is defined as
\begin{equation}
  \beta^2_\rho = k^2 - k^2_z,
\end{equation}
where
\begin{equation}
  k^2 = \omega^2\mu_0\mu_r\varepsilon_0\varepsilon_r 
\end{equation}
is free space wave number and $k_z$ is the longitudinal wavenumber,
which includes a complex resonance frequency $\omega_0=\omega_{0Re
}+i\omega_{0Im }$ at resonance ($\omega\to\omega_0$). The $Q$-factor
of the composed multi-concentric layered structure is computed as in
\cite{Eriksson_res_tunell} 
\begin{equation}
Q_0=\frac{\omega_{0Re}}{ 2\omega_{0Im}}.
\end{equation}
Setting the radial propagation constant
$\nu := \beta_\rho  =\sqrt{k^2-k²_z}$, and assuming that there is no
variation in z-direction (i.e. a pure radial resonance) renders the
propagation constant  as $\nu=\omega\sqrt{\mu\varepsilon}$
with $\mu \varepsilon=\mu_0\mu_r\varepsilon_0\varepsilon_r$.

Let us now move to establish a suitable radial impedance
transformation model. Since the wave impedance is anisotropic in
radial direction \cite{Harrington}, the reflection coefficient must be
derived accordingly.
Let  $Y^+_C$ and  $Y^-_C$ be the anisotropic admittances in outward
and inward directions, respectively. We define the reflection at the
load as $\Gamma=V^-/V^+$,  where $V^-, V^+$ are reflected and incident
voltage at load $Z_L$, respectively, the outgoing
current as $I^+=Y^+_CV^+$ and the reflected current as $I^-=Y^-_CV^-$.
Finally, $V^++V^-=V_L$, $I^+-I^-=I_L$ renders a reflection coefficient
at the load impedance $Z_L$:

\begin{equation}\label{eq:2}
  \Gamma_L=\frac{Z^-_C(Z_L-Z^+_C)}{Z^+_C(Z_L+Z^-_C)},
\end{equation}
where characteristic anisotropic impedance $Z^-_C=(Y^-_C)^{-1}$ and
$Z^+_C=(Y^+_C)^{-1}$.

It is noticed that equation (\ref{eq:2}) simplifies to: 
\begin{equation}
  \Gamma_L=\frac{Z_L-Z_C}{Z_L+Z_C}
\end{equation}
for isotropic characteristic impedance.

\subsubsection{\label{sec:imp_TE_wave}Impedance Transformation }

For notations used in this and following sections, we refer to table 2.
For cylindrical TE wave the characteristic impedance for a radial
outgoing TE wave \cite{Harrington} is given by the formula:
\begin{equation}
  Z_{Cout}=\frac{E_{\phi}}{H_{z}}=i\frac{\omega\mu H_m^{(2)\prime}(\nu
    r)}{\nu H_m^{(2)}(\nu r)}.
\end{equation}

The characteristic impedance for a radial incoming wave is given by
the formula: 
\begin{equation}
  Z_{Cin}=\frac{E_{\phi}}{H_{z}}=-i\frac{\omega\mu H_m^{(1)\prime}(\nu r)}{\nu H_m^{(1)}(\nu r)}.
\end{equation}
With the voltage ratio $\frac{V_0^-}{V_0^+}$ at the load at $r_0$ given by the formula:
\begin{equation}
  \frac{V_0^-}{V_0^+}=\frac{H_m^{(2)\prime}(\nu r)Z_{Cin}(r_0)(Z_L-Z_{Cout}(r_0))}
       {H_m^{(1)\prime}(\nu r)Z_{Cout}(r_0)(Z_L+Z_{Cin}(r_0))}
\end{equation}
the transformed load impedance from $r_0$ to $r$ becomes
\begin{equation}\label{zout1}
  Z_{out}(r) =
  \frac{Z_{Cin}(1+\Gamma(r))}{\frac{Z_{Cin}}{Z_{Cout}}-\Gamma(r)} = \\
  \frac{Z_{Cin}(1+\frac{V_0^-H_m^{(1)\prime}(\nu r)}{V_0^+H_m^{(2)\prime}(\nu r)})}
       {\frac{Z_{Cin}}{Z_{Cout}}-\frac{V_0^-H_m^{(1)\prime}(\nu r)}{V_0^+H_m^{(2)\prime}(\nu r)}}.
\end{equation}
We simplify last expression to get numerically efficient formulas  which are used in computations by introducing notations
\begin{equation}
\begin{split}
  F_1(x) &= Z_LH_m^{(2)}(x)-BH_m^{(2)\prime}(x), \\
  F_2(x) &= BH_m^{(1)\prime}(x)-Z_LH_m^{(1)}(x),  \\
\end{split}
\end{equation}
where the load
$Z_L$ is given at radius $r_0$ (for $r_0>r$). Then (\ref{zout1}) can be rewritten as
\begin{equation}
\begin{split}
  Z_{out}(r) &= 
  \frac{BH_m^{(1)\prime}(\nu r)F_1(\nu r_0) + H_m^{(2)\prime}(\nu r)F_2(\nu r_0)}
       {H_m^{(1)}(\nu r)F_1(\nu r_0) + H_m^{(2)}(\nu r)F_2(\nu r_0)},
\end{split}
\end{equation}
with $B=\frac{i\omega\mu}{\nu}=i\sqrt{\frac{\mu}{\varepsilon}}$. 
We scale this expression with exponential function in order to handle
finite metal conductivity: 

\begin{equation}
  Z_{out}(r) =    \frac{ B e^{2i\nu(r-r_0)}H_m^{(1)\prime}(\nu r)F_1(\nu r_0)
    + H_m^{(2)\prime}(\nu r)F_2(\nu r_0)}
       {e^{2i\nu(r-r_0)}H_m^{(1)}(\nu r)F_1(\nu r_0)
         + H_m^{(2)}(\nu r_0)F_2(\nu r_0)}.
\end{equation}

The inward input impedance into the innermost region is given by the
formula: 
\begin{equation}
  Z_{in}(r) = B \frac{J'_m(\nu r)}{J_m(\nu r)}.
\end{equation}
Resonance condition is fulfilled when
$Z_{in}(r_{bound})+Z_{out}(r_{bound}) = 0$ for both real and imaginary
parts. The  resonance condition can be calculated at any $r_{bound}$
inside the circular cylindrical region. In this work $r_{bound}=R_2$
(see Figure \ref{fig:cross_section}).

\subsubsection{\label{sec:imp_trans_TM_wave}Impedance Transformation
  For cylindrical TM wave}
We follow the same steps 
 as for TE wave, except that the characteristic
impedance for an outgoing wave is given by the formula:
\begin{equation}
  Z_{Cout} = -\frac{E_z}{H_\phi} =
  - \frac{i\nu H_m^{(2)}(\nu r)}{\varepsilon k H_m^{(2)\prime}(\nu r)}.
\end{equation}
The characteristic impedance for an incoming wave can be computed as
\begin{equation}
  Z_{Cin} = -\frac{E_z}{H_\phi} =
  \frac{i\nu H_m^{(1)}(\nu r)}{\varepsilon k H_m^{(1)\prime}(\nu r)}.
\end{equation}
Introducing notations
\begin{equation}
\begin{split}
  F_3(x) &= Z_LH_m^{(1)\prime}(x)-AH_m^{(1)}(x), \\
  F_4(x) &= Z_LH_m^{(2)\prime}(x)-AH_m^{(2)}(x), \\
\end{split}
\end{equation}
we scale the obtained expression for $Z_{out}$ with exponential function in order to
handle finite metal conductivity:
\begin{equation}
\begin{split}
  Z_{out}(r) &= 
  \frac{AH_m^{(2)}(\nu r)F_3(\nu r_0) - e^{2iv(r-r_0)}H_m^{(1)}(\nu r)F_4(\nu r_0)}
       {H_m^{(2)\prime}(\nu r)F_3(\nu r_0)
         - e^{2iv(r-r_0)}H_m^{(1)\prime}(\nu r)F_4(\nu r_0)},
\end{split}
\end{equation}
where $A=\frac{i\nu}{\varepsilon k}=i\sqrt{\frac{\mu}{\varepsilon}}$.

The inward input impedance into the innermost region can be computed
simply as:
\begin{equation}
  Z_{in}(r) = A\frac{J_m(\nu r)}{J'_m(\nu r)}.
\end{equation}

A test code (the same code as used for generating electromagnetic
fields in \cite{Beilina}) based on a spectral domain Greens function for
cylindrical geometry \cite{Sipus} was compared to the transverse
resonance method for both TE and TM modes for verification (with
$k_z = 0$: where an excitation current in z-direction renders TM
modes, and an excitation in transverse angular $\phi$ direction,
renders TE modes).

\subsubsection{\label{sec:TE110_TE210_1-dim}TE110 and TE210 Mode Field
  Distribution for 1-Dimensional Radial Tomography} 
In the case when angular index $m\!\gg\!1$, then the electric field is
dominant near resonator radius (see the equivalent field distribution
for parallel plate TM disc resonators \cite{Eriksson_mode_chart}). Even
TE210 mode has electric field significantly more confined near pipe
radius compared to TE110, which have a more homogeneous electric field
distribution. This can be exploited tomographically, since TE210 mode
field pattern penetrates less radially inwards than compared to the
TE110 mode. Thus, the TE210 mode is more sensitive to the presence of
an outer concentric dielectric layer than compared to the TE110 mode.

Having the quality factor and resonance frequency for each TE110 and
TE210 mode, a set of four unknown material and dimensional parameters
can in theory be extracted using the same transverse resonance
technique described previously.

\subsubsection{\label{sec:open_ended_coax}Open Ended Coax Quarter-wave
  Resonator Probe}

An open ended coax quarter-wave resonator exposed to pipe can serve as
an additional measurement probe. If dimensioned properly (i.e. with
suitable coax diameters $a$ and $b$), its penetration depth may be shorter
than even the TE210 mode. The open-ended coax resonator gives
typically a quality factor and resonance frequency for low-loss
exposed media ("open-circuit" type load) as well as high loss exposed
media ("short-circuit"). For the latter case, the frequency shift is
negligible, while amplitude changes decreases with increased electric
media loss. An intermediate region between low and high loss renders a
rather "arbitrary" wave-form, where the open end of coax sees more of
"matched load" impedance. Thus, in the low-loss region, the resonator
is "open-ended quarter wave", while in high-loss region, it is
half-wave type resonator (high electrical loss imply higher electrical
conductive load). 

In this work, the direct magnitude reflection response is minimized
with respect to the model of open ended coax and its quarter-wave
transmission-line circuitry - without any intermediate resonance
frequency and $Q$-factor calculations.

We apply the full-wave Hankel transform based model in
\cite{Baker_Jarvis}. If the pipe diameter is significantly larger than
the coax outer diameter $b$, the planar ground plane model in
\cite{Baker_Jarvis} is assumed still to be valid. It is noted that the
abrupt discontinuity from the coax section into a grounded plane
excites higher order terms apart from the fundamental incident coax
TEM mode. In \cite{Baker_Jarvis}, these higher modes are TM modes with
only radial variations due to the angular symmetry. If the coax
diameter $b$ increases relative to the pipe diameter $D$, the angular
variations of the basis functions/higher modes would be stronger -
however not estimated to be as large as the radial higher mode
excitations. 

A compromise in accuracy would be to assume an incident ideal TEM
wave, and matching the tangential electric and magnetic fields at the
curved open ended coax- pipe interface using a spectral domain
approach \cite{Sipus}, and using basis functions/modes with angular
dependency (in local open-ended coax coordinate system). 

For suitable coax dimensions $a$, $b$, a third characteristic
penetration depth (smaller than TE210) can be obtained, and thus, one
can replace one of the four transverse resonance equations with an
equation for the open-ended coax resonator.

\begin{figure}
 \begin{center}
 \begin{tabular}{cc}
  \includegraphics[width= 0.5 \linewidth ]{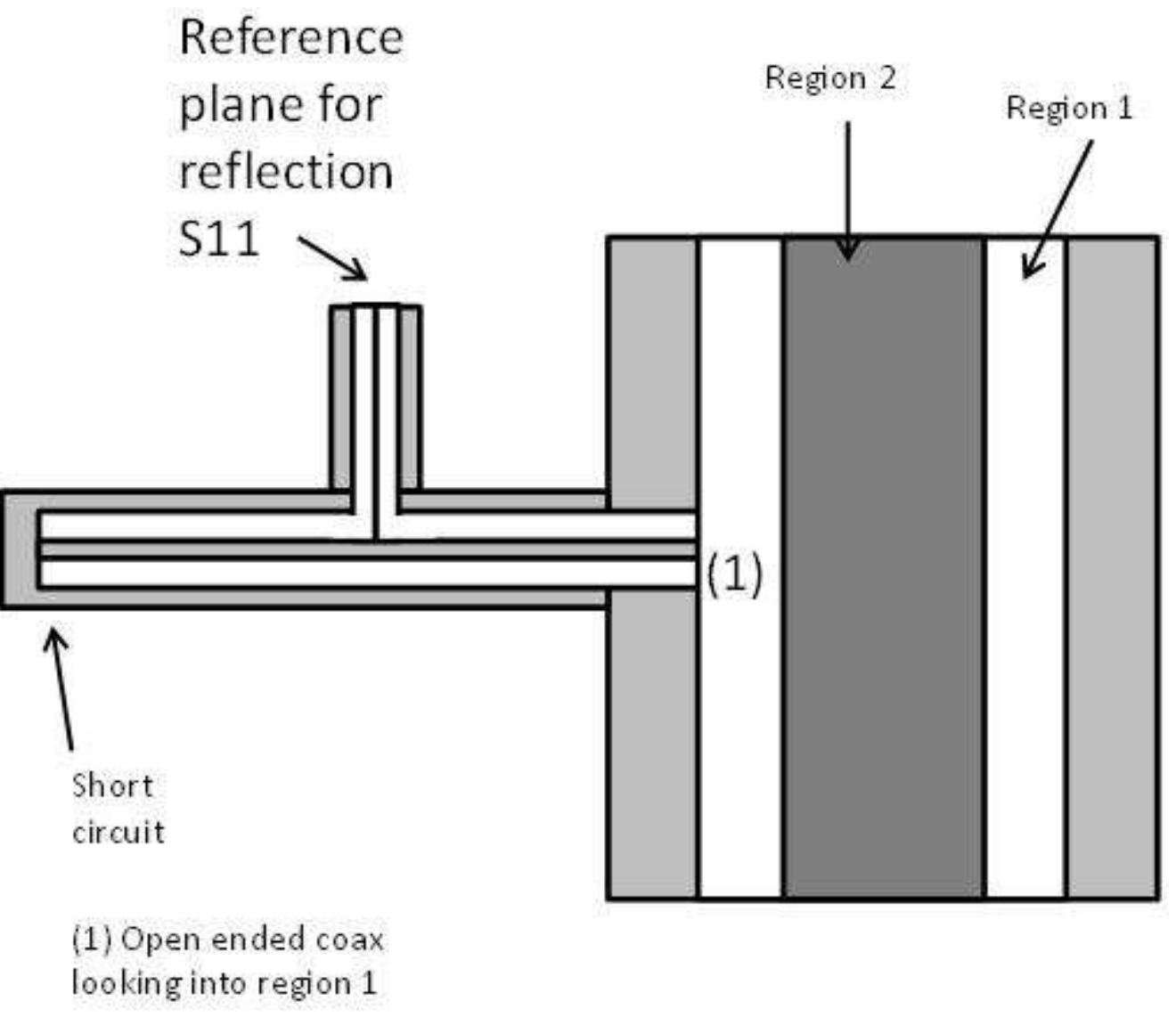} &
  \includegraphics[width= 0.5 \linewidth]{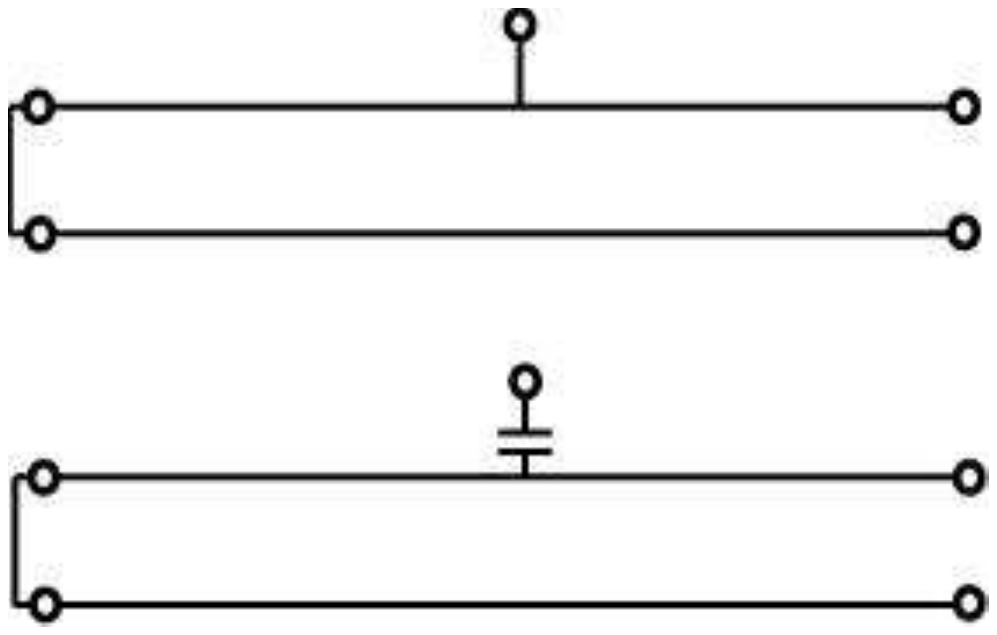}
 \end{tabular}
 \end{center}
  \caption{At the left is open ended galvanic coupled quarter-wave
    coaxial resonator connected flush to waveguide pipe section. At
    the  right is schematic circuits of galvanic and capacitive
    coupled quarter-wave resonators} 
  \label{fig:quarter_wave}
\end{figure}

As model for input impedance $Z_{liq}$,  a full-wave
model of \cite{Baker_Jarvis} is employed; using fundamental TEM mode, and two lowest TM modes
as basis functions, and the input impedance is calculated from the
obtained TEM reflection coefficient. The reflection coefficient
$S_{11}$ at reference plane in figure \ref{fig:quarter_wave}(a) is
obtained by standard impedance transformations.

Typically, for low-salinity exposed media, the response in $S_{11}$ are
resonance poles shifting down in frequency with increased water
content of the exposed media. For a "high" saline media, the
quarter-wave resonator turns in effect into a half-wave resonator due
to the high conductive saline media that is exposed to the open coax
end. In this high salinity water continuous regime, the frequency
shift is very weak, while amplitude changes are still significant -
even for salinities above 12\%, but the amplitude change starts to
decrease with increasing salinity in this high salinity region. 

\subsubsection{\label{sec:quarter_coax_coupling}Coupling to the
  open-ended quarter wave coax resonator}
By probing the resonator in the middle of the coax instead of at the
left end, separation of the resonance frequencies  is narrowed for the
same physical resonator length.
Practically (if not also theoretically) is impossible to select an
optimal coupling to a quarter-wave open ended coax loaded with media,
while having a large range for the imaginary part as well as real part
(with salinities from 0 up to 25\% the imaginary part of saline water
changes with several orders of magnitude). The magnitude of the
response ($mag(S_{11})$) will not have a monotonic change in amplitude
as permittivity real and imaginary parts change.

Experimentally we have  found  that using a pair of resonators - see lower figure
\ref{fig:quarter_wave} (one that is simply galvanic coupled, and the another one is capacitive coupled, in this work a coupling capacitance of 10 pF), a better sensitivity
in amplitude and frequency shift is achieved.
We should note that a pair of capacitive and galvanic coupled
resonators cannot render more information than a phase and magnitude
reflection measurement directly at the open end of the coax, but
rather, they (resonator pair) transform the complex reflection data to
resonance type response, so that the benefit of both amplitude change
as well as frequency shift can be taken advantage of.

\subsection{\label{sec:}Reconstruction}

Using the models presented in previous sections, an algorithm can be
built that reconstructs the distribution of media from the measured
spectra. It is noticed that the 4 transverse resonance equations are
minimized in the same manner regardless parameters. Since there are 4
equations at hand and 4 unknowns, it is suitable to apply Newton's
method.

The full computational scheme consists of 2 steps, see figure
\ref{fig:peak_fit} for result of the first step.  First the resonance
frequency and loaded quality factors are extracted for TE110 and TE210
modes. Typically, measured data is a transmission measurement - with
weak coupling, the unloaded $Q$-factor can be approximated by the
measured loaded $Q$ - or otherwise, a more sophisticated transfer
function that models coupling circuitry as well must be applied to
obtain the unloaded $Q$ and resonance frequency.

\begin{figure}[ht]

  \includegraphics[width=0.5\linewidth]{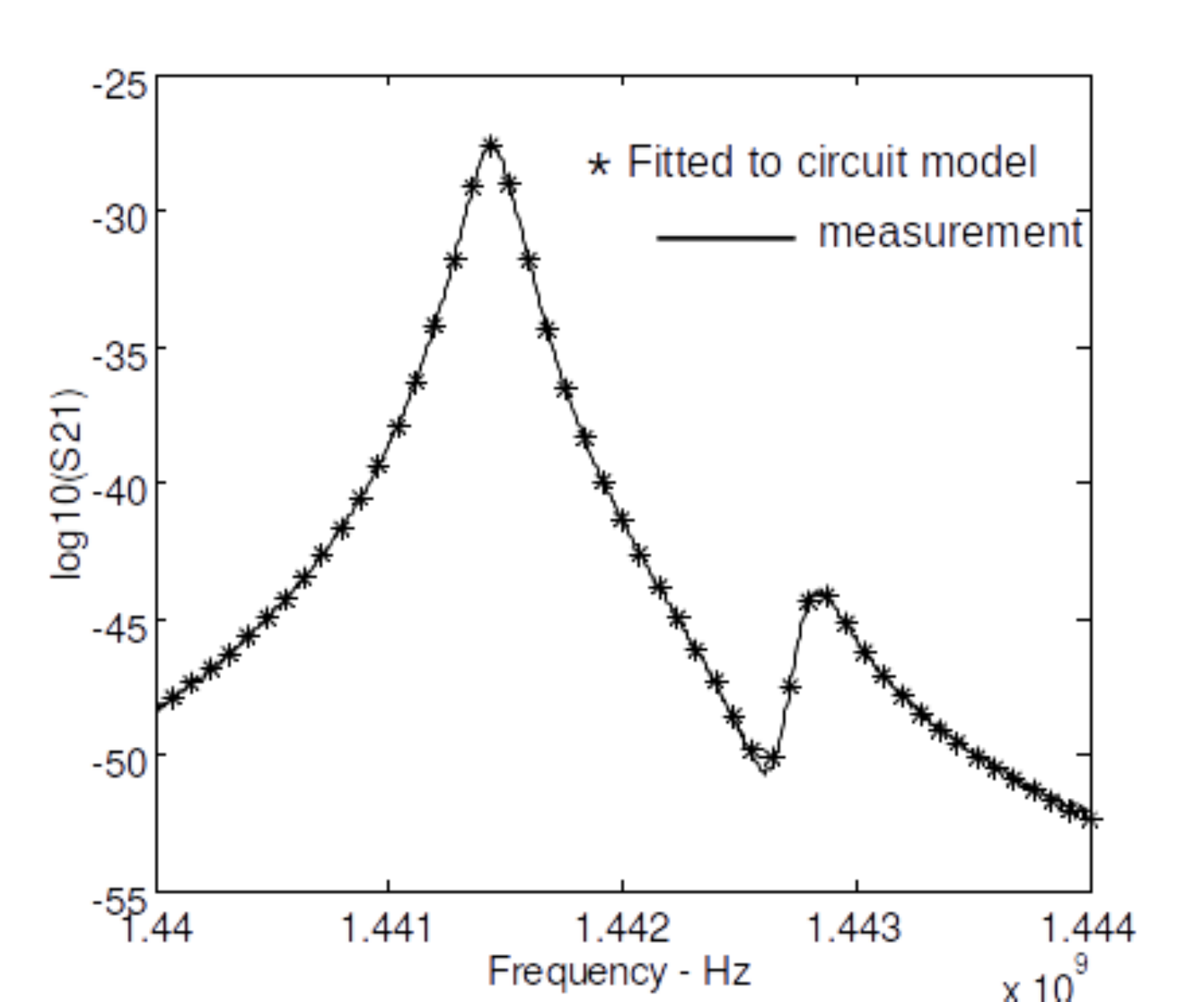}
  \includegraphics[width=0.5\linewidth]{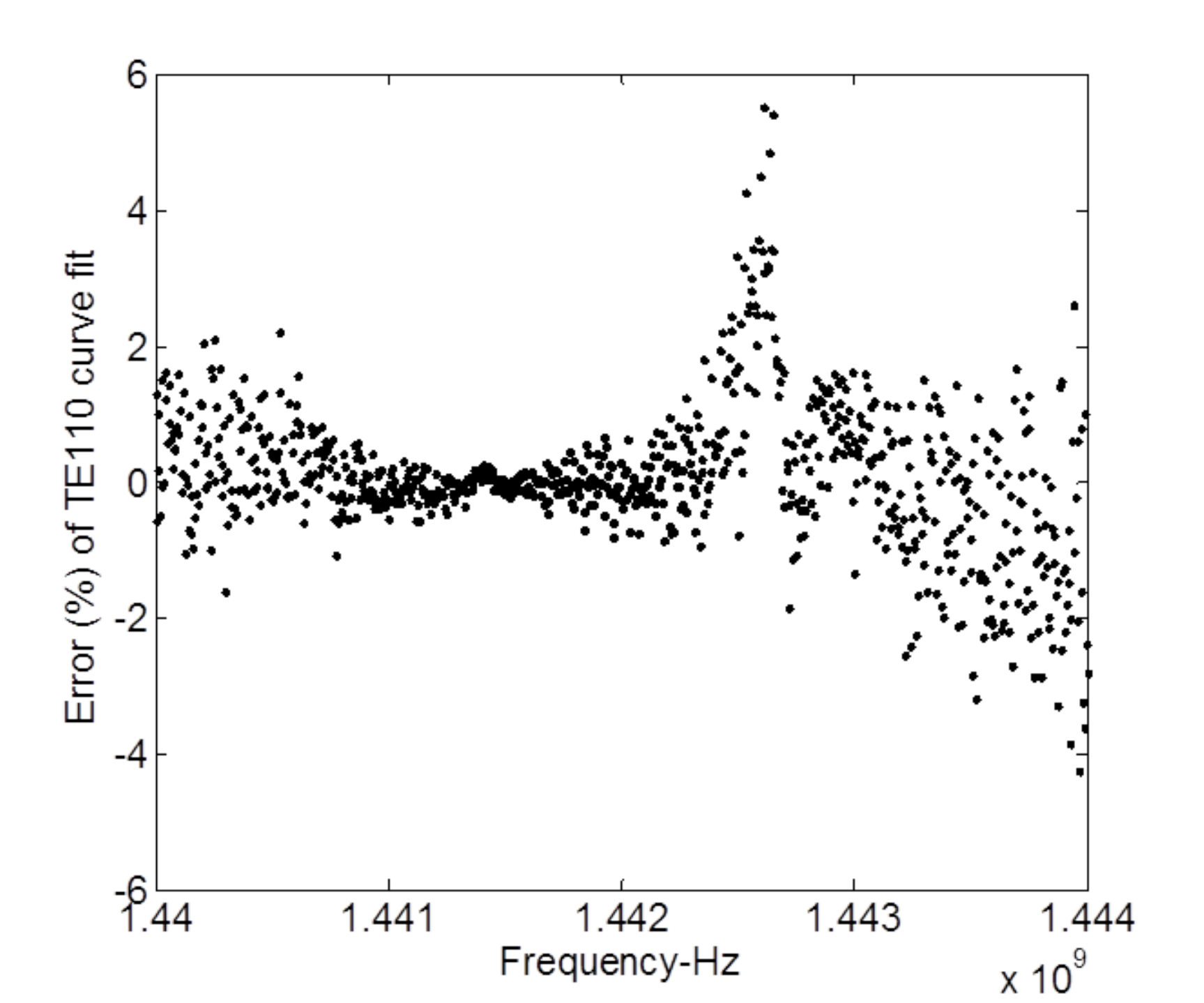}
  \includegraphics[width=0.5\linewidth]{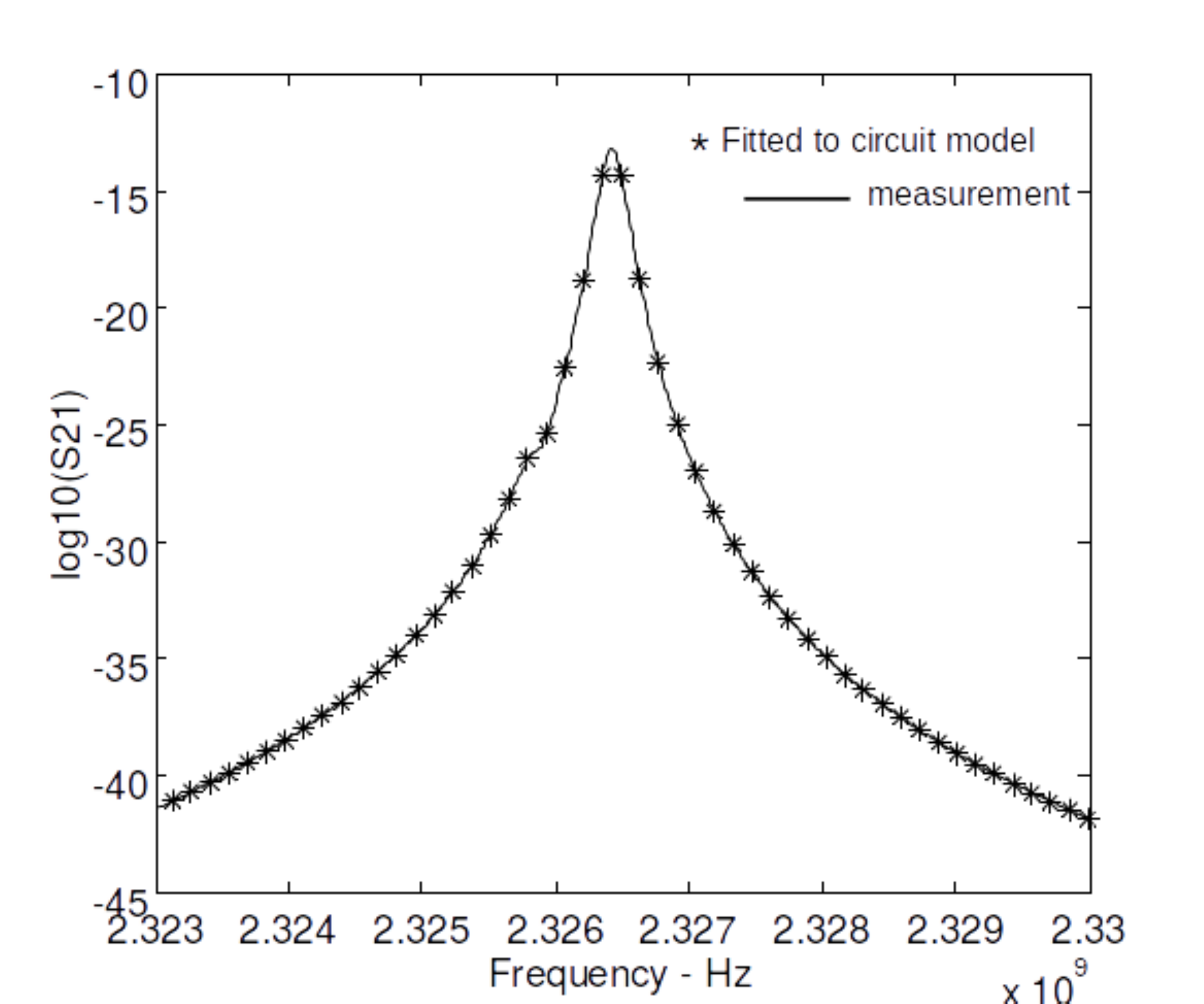}
  \includegraphics[width=0.5\linewidth]{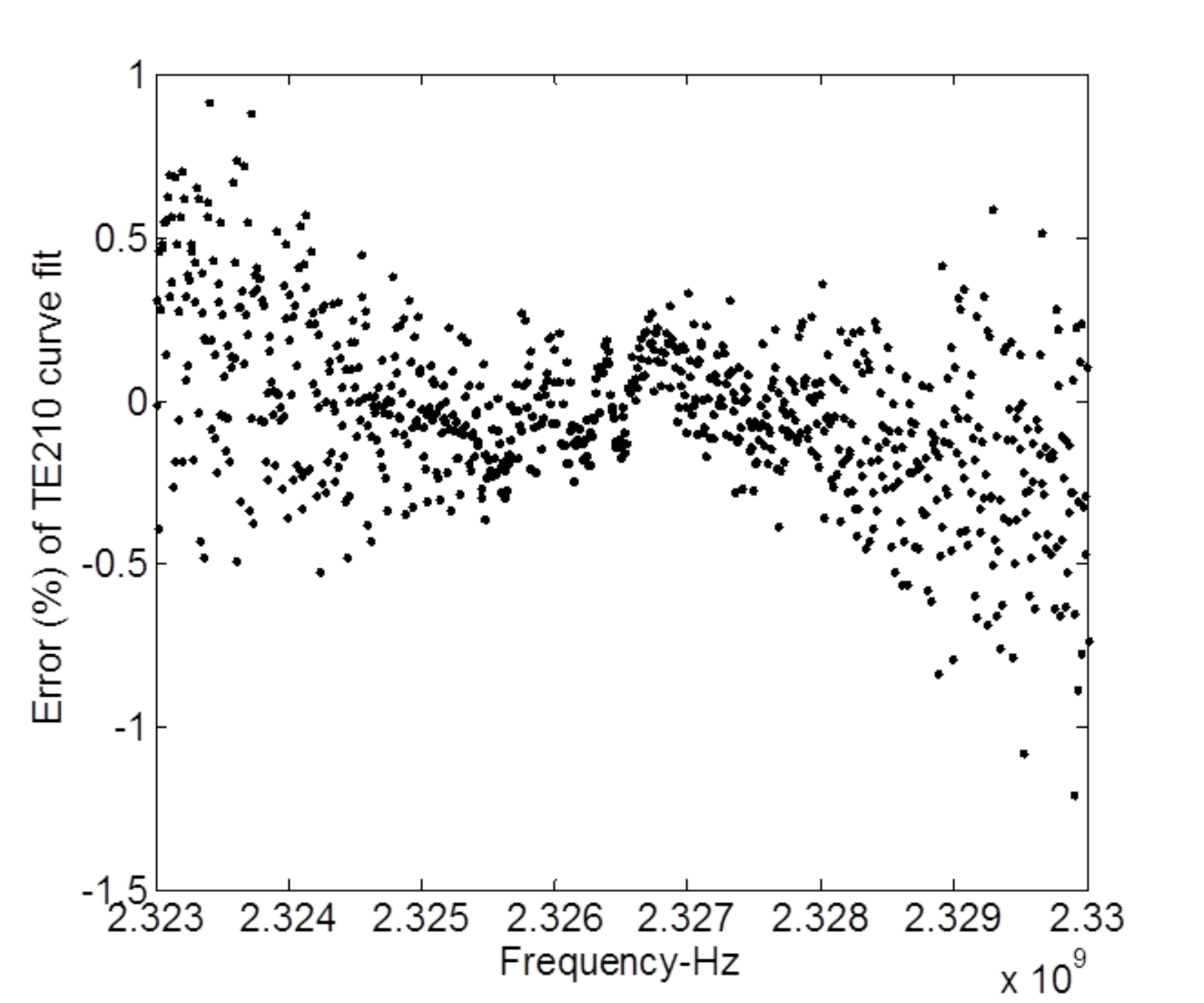}
  \caption{{Result of step 1. Behavior of the
      Lorenzian function as a function of frequency, fitted to
      experimentally measured TE110 (top figure) and a TE210 (bottom
      figure) resonance in an air filled aluminium pipe of 128.5mm
      internal diameter. Resonant frequencies and quality factors are
      extracted from a Lorentzian fit the transmission
      measurement. Complex resonant frequencies are calculated as
      $f_0^{TEm10} + i\frac{f_0^{TEm10}}{2Q_0^{TEm10}}$, which form
      input  to the combined transverse resonant open ended coax 
      functional. The figures on the right side show the relative
      difference between the measured data and the Lorenzian function.}} 
  \label{fig:peak_fit}
\end{figure}

Using results of work \cite{Petersan} we can conclude that for
resonance spectrum data with signal-to-noise ratio $<65$, a non-linear
least squares fit to a Lorentzian curve is more accurate than fit to
the phase vs. frequency. Thus, for a sufficiently weakly coupled
resonator under test, where unloaded $Q$ can be closer in value to the
measured loaded $Q$, the transmitted resonance spectrum is anticipated
to have more noise simply due to the relatively low signal level. One
may then conclude that non-linear least squares fit to a Lorentzian
curve is suitable for characterizing weakly coupled resonator
configurations.  We refer to \cite{Petersan} for review of different
methods for caclculating of resonance spectrum, extraction of
$Q$-factor and resonance frequency.


The second step of our computational procedure is to use the measured
$Q$-factors and resonant frequencies for TE110 and TE210 modes as
input to a combined transverse resonance condition for TE110 and TE210
modes, and solve then the equations using the optimization algorithm.
In this work the Lorentzian function $L(\omega)$ is defined by the
equalant circuit for the transverse resonator as shown in figure
\ref{fig:equiv_circuit} and it is given by the formula
\begin{equation}
  L(\omega) = A\cdot\left((\frac{\omega_0}{2Q_0})^2 \left((\omega-\omega_0)^2+(\frac{\omega_0}{2Q_0})^2\right)\right)+B,
\end{equation}
 where $\omega=2\pi f$ and
$\omega_0=2\pi f_0$, $f_0$ is resonance frequency and $Q_0$ is quality
factor, $A$ is amplitude and $B$ is constant.
 It is noticed that a 
realistic resonance curve has parasitic contributions from coupling
circuitry, or parasitic parallel capacitance. Also, probes/antennas
for resonance excitation may have intrinsic resonances which are
"multiplied" to the resonance of interest.

\begin{figure}[ht]
  \includegraphics[width=\linewidth]{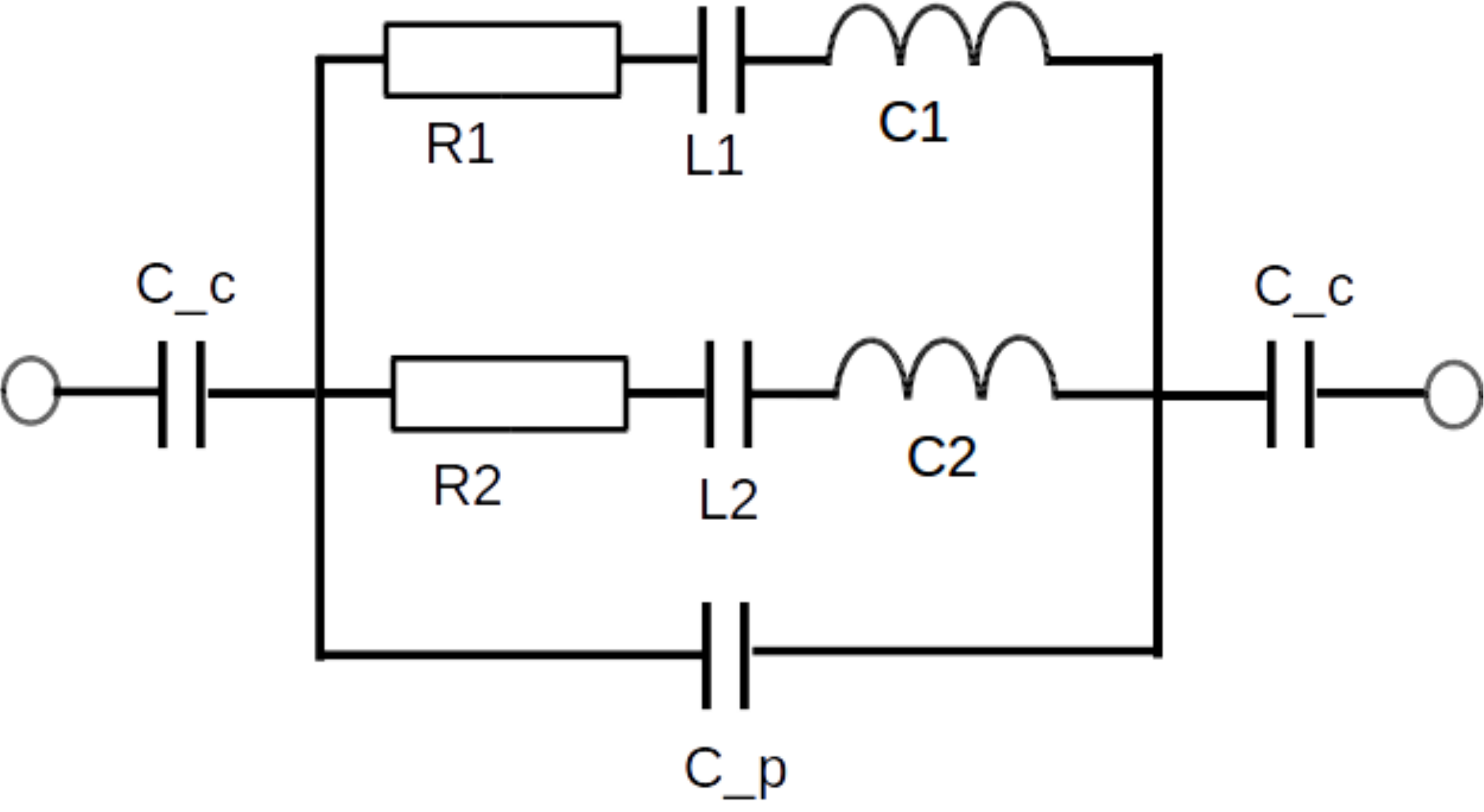}
  \caption{The diagram shows the equivalent electric circuit for the
    transverse resonator.}
  \label{fig:equiv_circuit}
\end{figure}

It is noticed that a 
realistic resonance curve has parasitic contributions from coupling
circuitry, or parasitic parallel capacitance. Also, probes/antennas
for resonance excitation may have intrinsic resonances which are
"multiplied" to the resonance of interest.

Typically, resonance frequency $f_0$ and quality-factor $Q$ can be
extracted by a simple peak search and a numerical direct extraction of
the quality-factor from a ratio of band-width at half maximum peak
value and resonance frequency. These extracted values are good initial 
guesses for $f_{init}^{\mathsf{TE}110}$, $Q_{init}^{\mathsf{TE}110}$ and
$f_{init}^{\mathsf{TE}210}$, $Q_{init}^{\mathsf{TE}110}$.

For one TE resonance, assuming that above conditions hold, we have
four unknowns to be extracted: amplitude $A_0^{\mathsf{TE}110}$,
$Q_0^{\mathsf{TE}110}$, $f_0^{\mathsf{TE}110}$ and $B_0^{\mathsf{TE}110}$.

The same is valid for TE210 resonance. Only $Q_0^{\mathsf{TE}110}$,
$f_0^{\mathsf{TE}110}$ and $Q_0^{\mathsf{TE}210}$,
$f_0^{\mathsf{TE}210}$ are needed in  order to extract the four
unknowns liquid thickness, salinity, WLR and droplet ratio.

The measured quality factors $Q_0^{\mathsf{TE}110}$ and
$Q_0^{\mathsf{TE}210}$ as well as the measured resonance frequencies
$f_0^{\mathsf{TE}110}$ and $f_0^{\mathsf{TE}210}$ can be approximated
to the unloaded corresponding entities as long as the coupling to the
resonator is sufficiently weak - otherwise, the unloaded quality
factors and resonant frequencies must be calculated by an analysis
that includes the coupling circuitry influence. Here, for simplicity,
we assume that the coupling is weak and that the unloaded and measured
loaded entities are the same.

The reconstruction of the media content and distribution is solved by
minimizing squared real and imaginary parts of transverse resonance
functionals for TE110 and TE210:
{\small
\begin{equation} 
\begin{split}
  J^{Re}_{W}(f_0^{W},Q_0^{W}, x, K) &= 
  \left(Re \left(Z_{in}(f_0^{W},Q_0^{W}, x, K) +
  Z_{out}(f_0^{W},Q_0^{W}, x, K)\right)\right)^2
  = 0, \\
  J^{Im}_{W}(f_0^{W},Q_0^{W}, x, K) &= 
  \left(Im \left(Z_{in}(f_0^{W},Q_0^{W}, x, K) +
  Z_{out}(f_0^{W},Q_0^{W}, x, K)\right)\right)^2
  = 0
\end{split}
\end{equation}
}

\noindent
where $W$ is TE110 or TE210 mode, $x$ is the vector of 4 unknowns
which are $x = (h, R_{WLR}, s, R_{DGR})$, and $K$ is the static input
parameters (temperature, pipe inner diameter and hydrocarbon
permittivity).

Having four unknowns and four equations, Newton-Raphson method is
suitable, since only first order derivatives needs to be calculated in
a Jacobian matrix. One may add more sensors (the open ended coax
resonator sensor shown in Fig. \ref{fig:quarter_wave}, for instance) so
that an overdetermined non-linear  system of equations is obtained. This
overdetermined system may be reduced back to a set of 4 equations
either by adding the sensor functionals $J_{quarter}$ (in
eq. \ref{eq20}) to the existing 4 transverse resonance functionals. It
is also possible to replace one of the existing transverse resonance
functionals with the sensor functional $J_{quarter}$. To obtain 4
unknown parameters $(h, R_{WLR}, s, R_{DGR})$ we minimize the
difference of the model value for reflection coefficient $S^1_{11}$
(in magnitude) and the magnitude of the measured reflection
coefficient $\tilde{S}^1_{11}$.

The measurement is taken exacly at the same point (reference plane)
as the excitation (incident wave). A continuous wave in a certain
frequency range is used, rather than a pulse. In addition to the 4
unknowns, the functional $J_{\textit{quarter}}$ also depends on the
geometric dimensions and 
the dielectric media inside the pipe, in addition to the 4 unknowns.
Thus, our goal is to minimize the following functional:

{\small
\begin{equation}\label{eq20}
\begin{split}
  &J_{\textit{quarter}}(h, R_{WLR}, s, R_{DGR}) = \\
 & \frac{1}{2} \int^{\omega_2}_{\omega_1} 
\left(|S_{11}^1(h, R_{WLR}, s, R_{DGR}, a,  b, \omega')| 
- |\tilde{S}^1_{11}(\omega')| \right)^2 d\omega' \\
&+   \frac{1}{2}\alpha_1\left(h-h_0\right)^2+\frac{1}{2}
  \alpha_2\left(R_{WLR}-R_{WLR_0}\right)^2 \\
&+   \frac{1}{2}\alpha_3\left(s-s_0\right)^2+\frac{1}{2}
  \alpha_4 \left(R_{DGR}-R_{DGR_0}\right)^2,
\end{split}
\end{equation}

\noindent
where $\alpha_j, j=1,2,3,4$ are small regularization parameters, such that
$\alpha_j \in (0,1)$. They can be
chosen as constant values depending on the noise level $\delta$, or
iteratively using one of the iterative regularization algorithms, see
\cite{TihonovYagola, BakushKokurinSmirnova} for some
of these algorithms. One of possible iterative choices for the
computing of regularization parameters (see \cite{BakushKokurinSmirnova,
  Samar_Masterthesis} for computational details) is
$\alpha^n_j=\alpha^0_j(n+1)^{-p}$, where $n$ is the number of iteration
in any gradient-like method (in our case  - number of iteration in
Newton's method), $p \in (0,1)$  and $\alpha^0_j$ are initial guesses for  $\alpha_j, j=1,...,4$. Similarly with
\cite{Klibanov_Bakushinsky_Beilina} we choose
$\alpha_j=\delta^{\gamma}$, where $\delta$ is the noise level and
$\gamma$ is a small number taken in the interval $(0,1)$.

The expression (\ref{eq20}) is in practice a sum due to the measured discrete
frequency points $\omega_i$ with steps $ \delta \omega'$: 
{\small
\begin{equation}
\begin{split}
  &J_{\textit{quarter}}(h, R_{WLR}, s, R_{DGR}) = \\
  &\sum^{N}_{i=0} \left(|S_{11}^1(h, R_{WLR}, s, R_{DGR}, a,
  b, \omega_i')| -  |\tilde{S}^1_{11}(\omega_i')| \right)^2 \cdot \delta \omega' \\
&+   \frac{1}{2}\alpha_1\left(h-h_0\right)^2+\frac{1}{2}
  \alpha_2\left(R_{WLR}-R_{WLR_0}\right)^2 \\
&+ 
  \frac{1}{2}\alpha_3\left(s-s_0\right)^2+\frac{1}{2}
  \alpha_4 \left(R_{DGR}-R_{DGR_0}\right)^2.
\end{split}
\end{equation}}

The resulting functional for two identical open ended coax
quarter-wave resonators - one capacitive and the second galvanic
coupled - is:

{\small
\begin{equation}
\begin{split}
 & J_{\textit{quarter}}(h, R_{WLR}, s, R_{DGR}) = \\
 & \sum^{N}_{i=0} \left(|S_{11}^{galv}(h, R_{WLR}, s, R_{DGR}, a,
  b, \omega_i')| - |\tilde{S}^{galv}_{11}(\omega_i')| \right)^2 \delta \omega'  \\
  &+\sum^{N}_{i=0} \left(|S_{11}^{cap}(h, R_{WLR}, s, R_{DGR}, a,
  b, \omega_i')| - |\tilde{S}^{cap}_{11}(\omega_i')| \right)^2  \delta \omega'  \\
 &+ \frac{1}{2}\alpha_1\left(h-h_0\right)^2+\frac{1}{2}
  \alpha_2 \left(R_{WLR}-R_{WLR_0}\right)^2 \\
  &+\frac{1}{2}\alpha_3 \left(s-s_0\right)^2+\frac{1}{2}
  \alpha_4 \left(R_{DGR}-R_{DGR_0}\right)^2.
\end{split}
\end{equation}}

\section{\label{sec:results}Results of simulation}

A set of simulations with different salinities, DGR, WLR and liquid
thickness $h$ was performed. In all cases, initial guesses were set to
the following:
\begin{align*}
s_0 :=  s^{init}&      = s^{true}\cdot 1.3, \\
R_{DGR_0} :=  R_{DGR}^{init}& = R_{DGR}^{true}\cdot 0.7, \\
R_{WLR_0} :=   R_{WLR}^{init}& = R_{WLR}^{true}\cdot 0.7, \\
 h_0 :=  R_h^{init}&    = h^{true}\cdot 1.3.
\end{align*}

The frequency range for the open ended coax resonator pair was adapted
in order to keep the resonance dips within a sufficiently wide
frequency range. No noise was added to the synthetic ``measured'' data. 
The conclusion is that the open ended coax resonator pair have
potential to significantly improve convergence and reduce the error
(defined as $ e = |x_{iterated}-x_{true}|/|x_{true}|$) by a factor of 100.

\begin{figure}[!t]
\begin{center}
  \includegraphics[width=\linewidth]{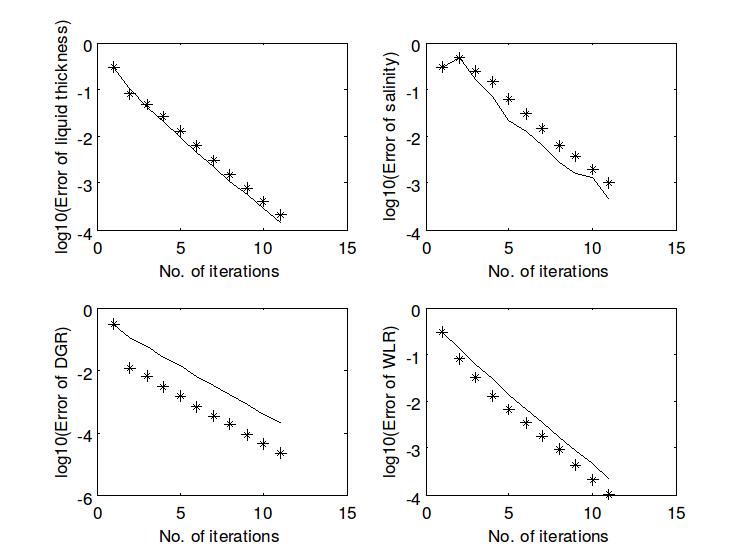}
  \includegraphics[width= 0.7\linewidth]{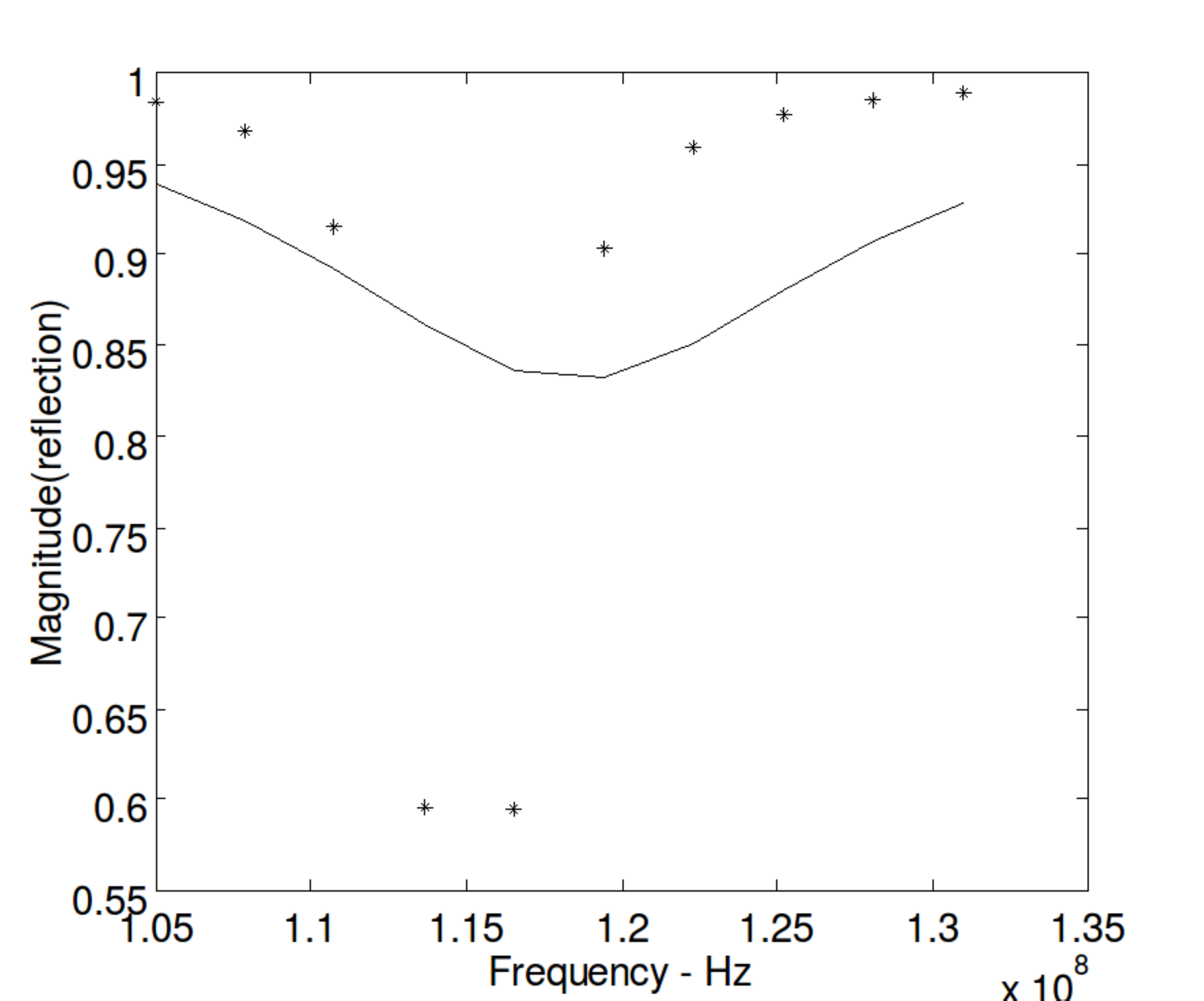}
\end{center}
  \caption{(Top) Error plots of simulation case for
    $s^{true}=10^{-4}$, $h^{true}=1$mm, $R_{DGR}^{true}=10^{-2}$,
    $R_{WLR}^{true}=0.1$. Solid line is without open ended quarter
    resonator pair. (Bottom) Magnitude of reflection of capacitive (+)
    and galvanic coupled (solid line) open ended coax resonators.}
  \label{fig:error_1}
\end{figure}

There is room for some optimization regarding the open ended coax
resonator pair -- for example the coupling capacitance value could be
further optimized. Number of frequency points and frequency range are
other details that may increase the benefits of having the open ended
coax resonator pair. Also, for some combinations of salinity, WLR, DGR
and liquid layer thickness, it may be more beneficial to only include
either capacitive or galvanic coupled open ended coax resonator to
increase convergence.

\begin{figure}[!t]
\begin{center}
  \includegraphics[width=\linewidth]{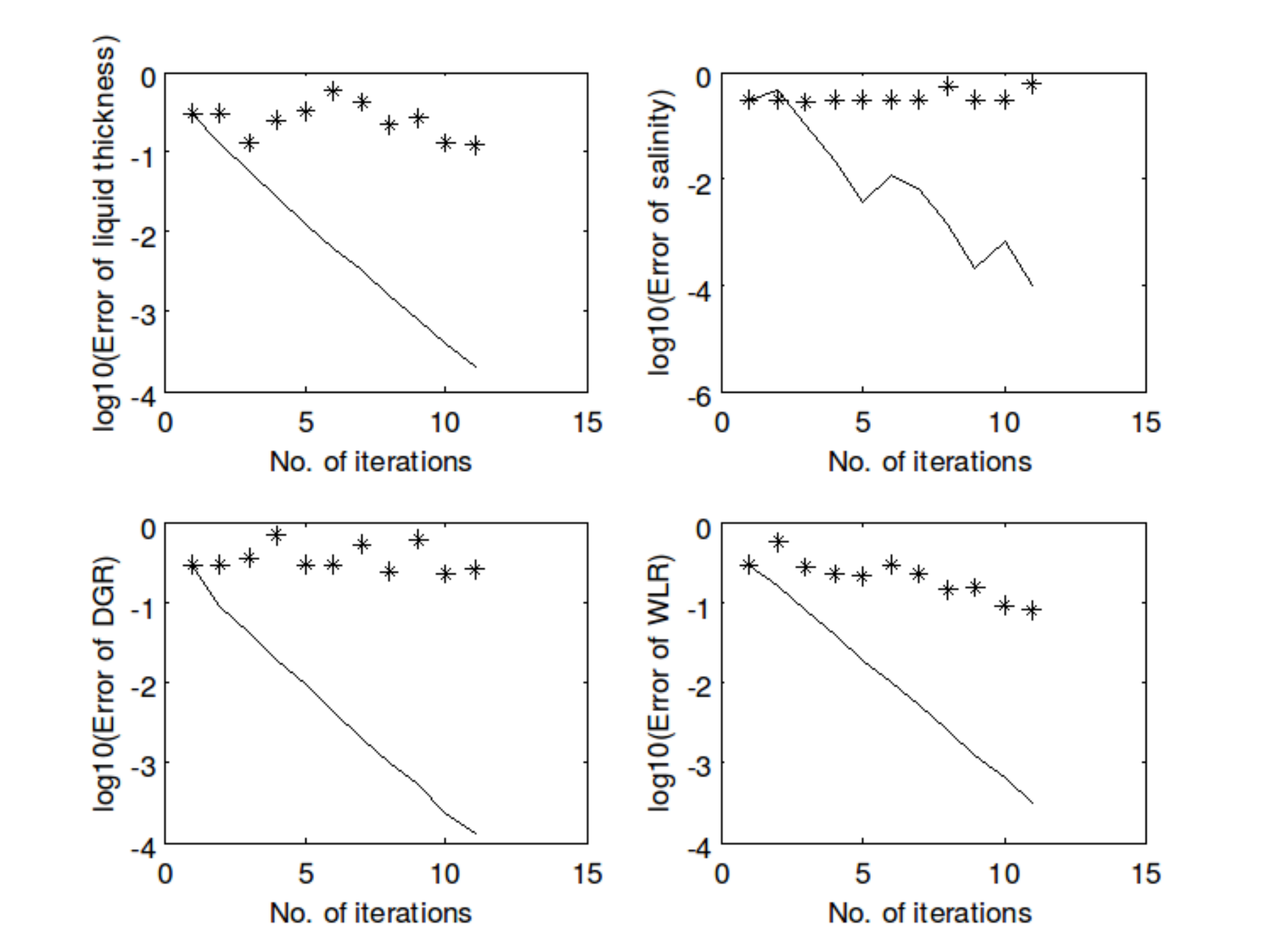}
  \includegraphics[width= 0.7\linewidth]{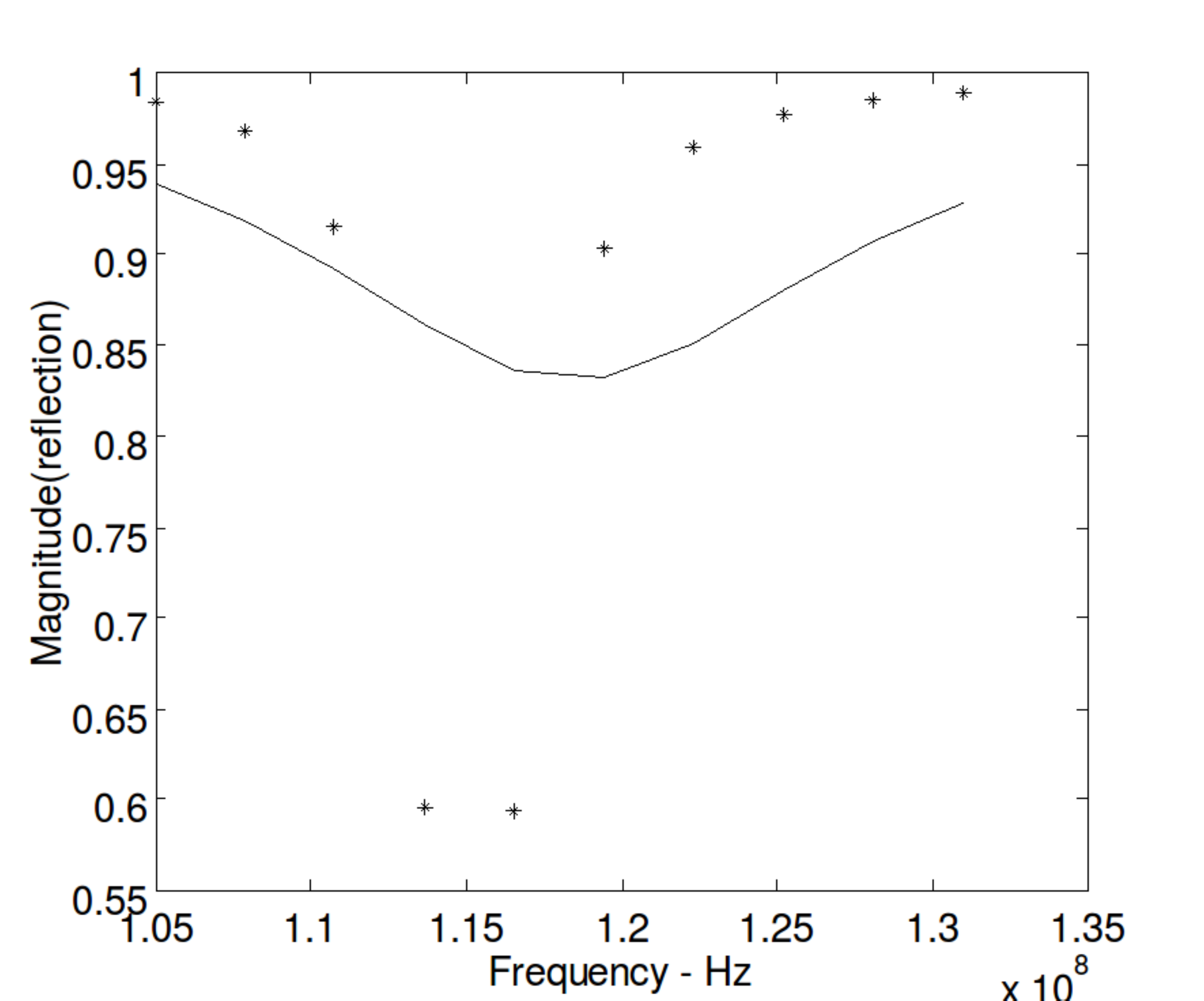}
\end{center}
  \caption{(Top) Error plots of simulation case for $s^{true}=10^{-4}$,
    $h^{true}=1$mm, $R_{DGR}^{true}=2\cdot10^{-3}$,
    $R_{WLR}^{true}=0.1$. Solid line is without open ended quarter
    resonator pair. (Bottom) Magnitude of reflection of capacitive (+)
    and galvanic coupled (solid line) open ended coax resonators.}
  \label{fig:error_2}
\end{figure}

\begin{figure}[ht]
\begin{center}
  \includegraphics[width=\linewidth]{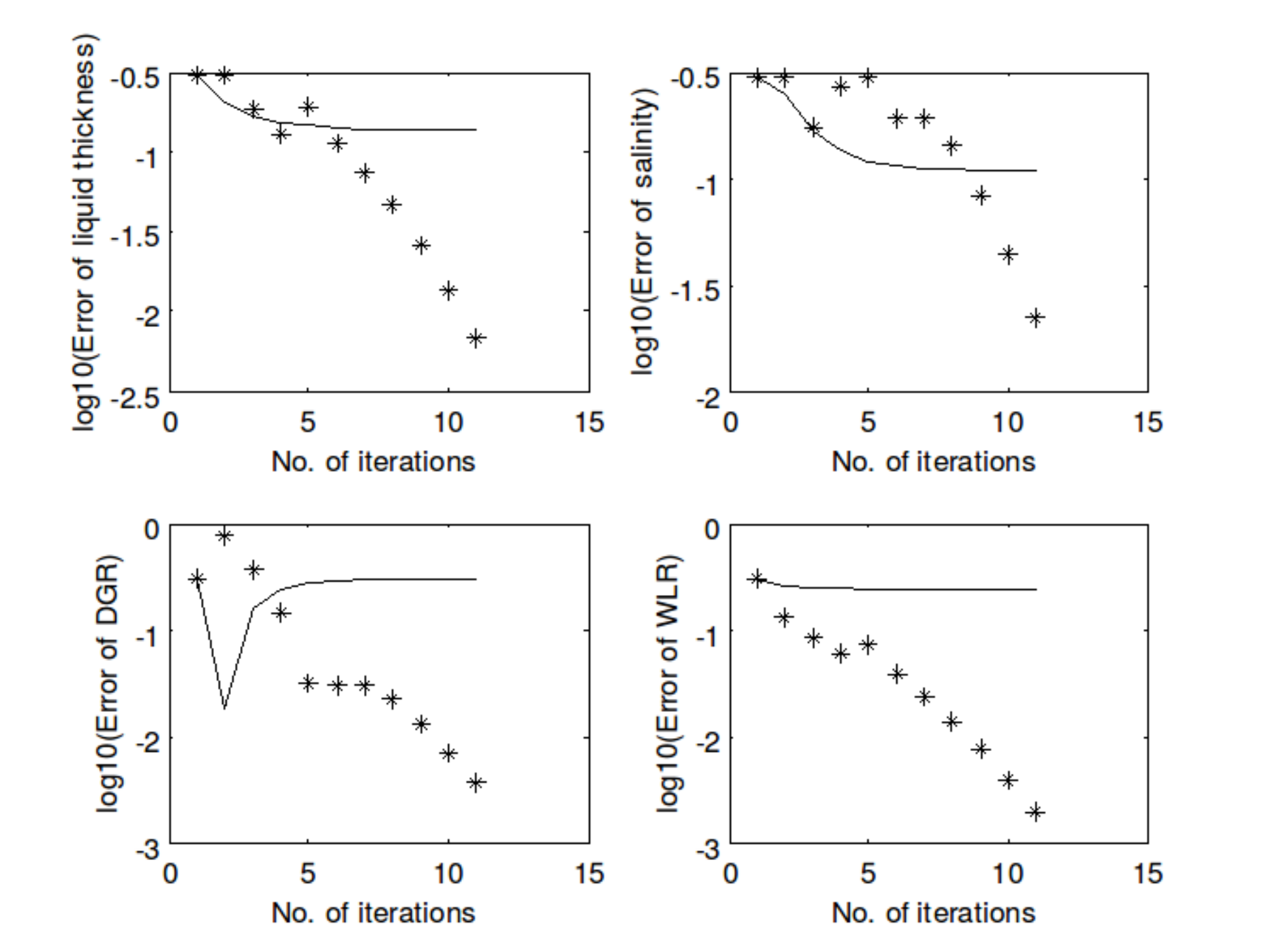}
  \includegraphics[width= 0.7\linewidth]{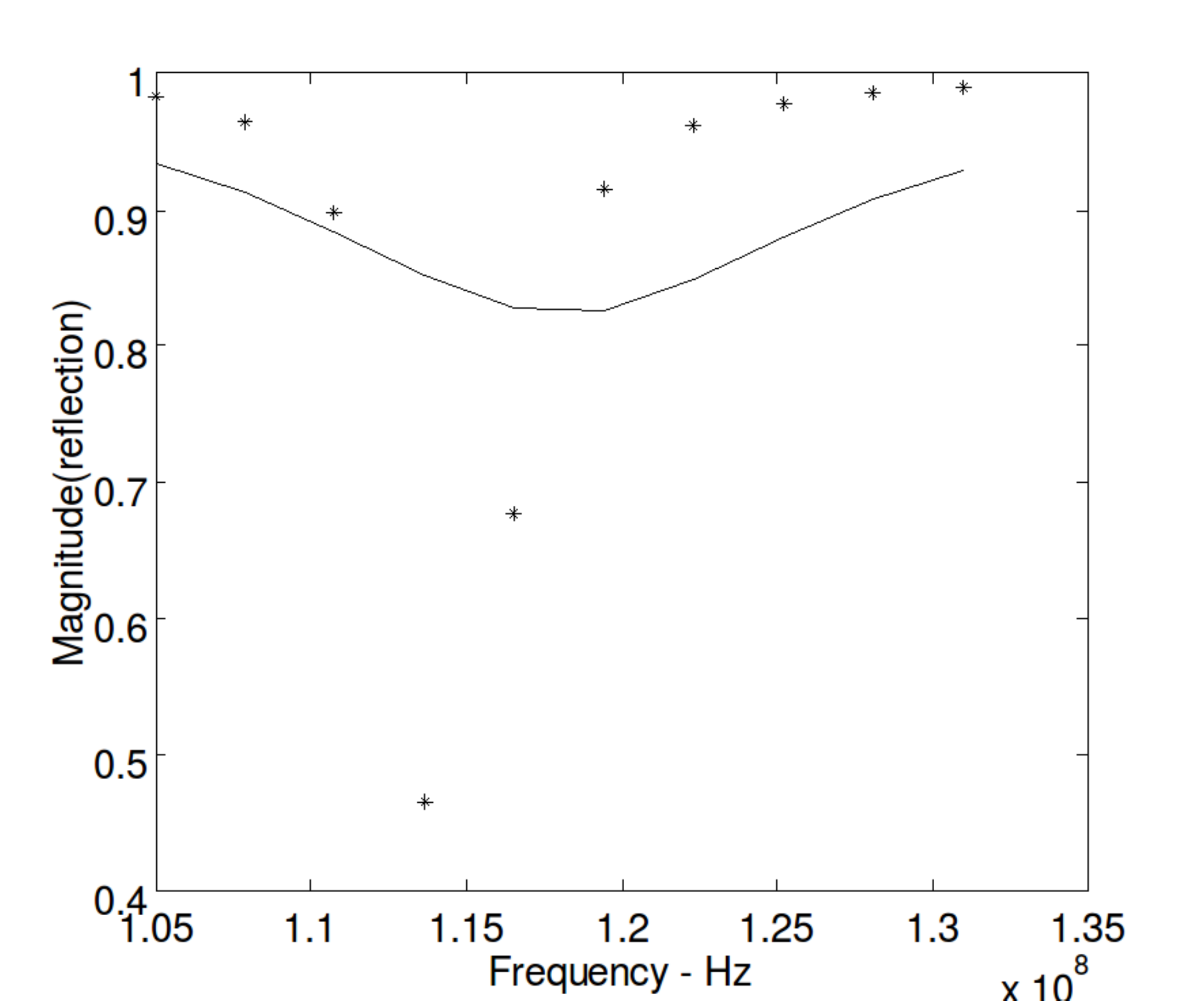}
\end{center}
  \caption{(Top) Error plots of simulation case for
    $s^{true}=10^{-4}$, $h^{true}=1$mm, $R_{DGR}^{true}=10^{-2}$, 
    $R_{WLR}^{true}=0.4$. Solid line is without open ended quarter
    resonator pair. (Bottom) Magnitude of reflection of capacitive (+)
    and galvanic coupled (solid line) open ended coax resonators.}
  \label{fig:error_3}
\end{figure}

\begin{figure}[ht]
\begin{center}
  \includegraphics[width=\linewidth]{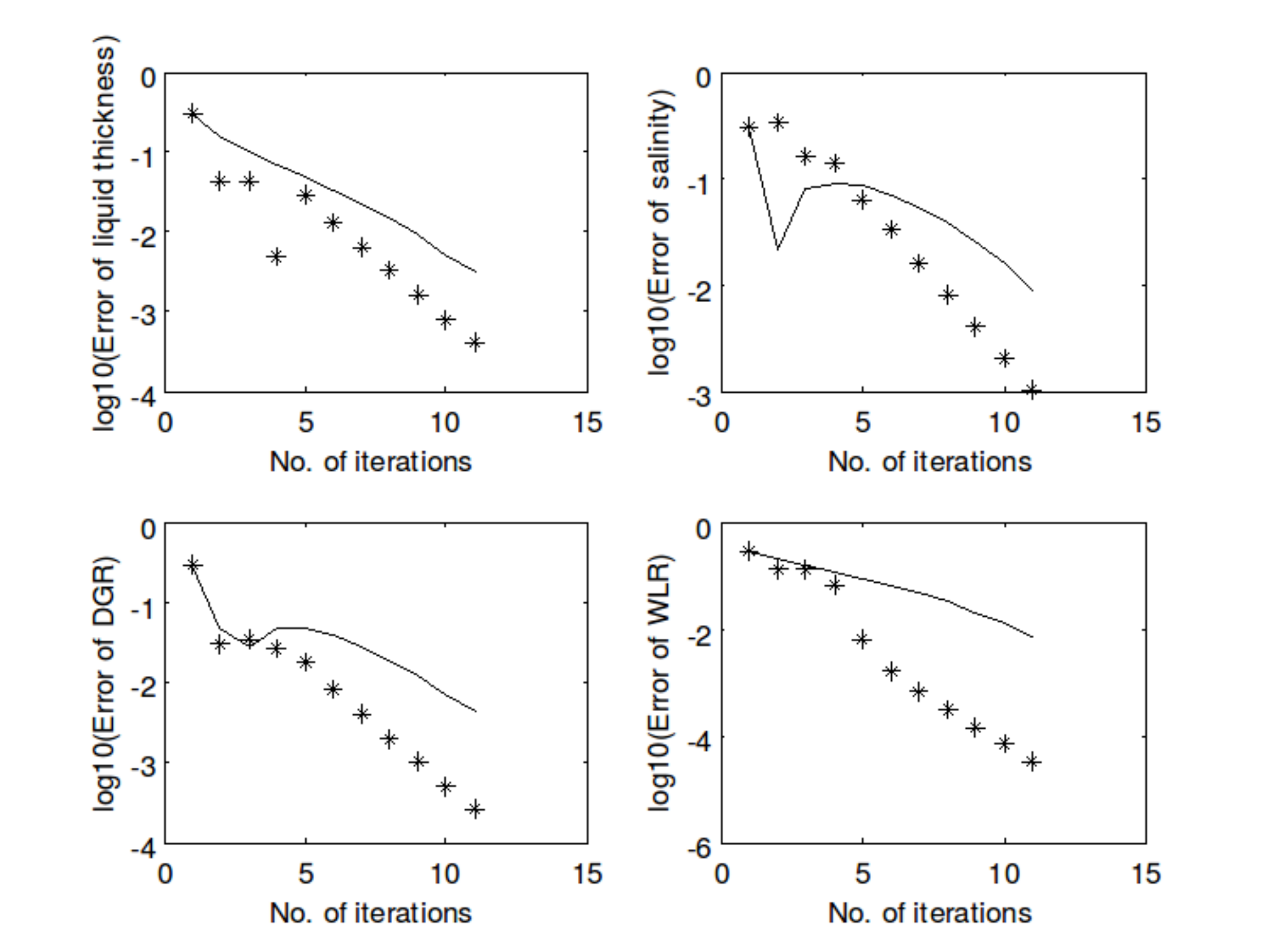}
  \includegraphics[width=0.7\linewidth]{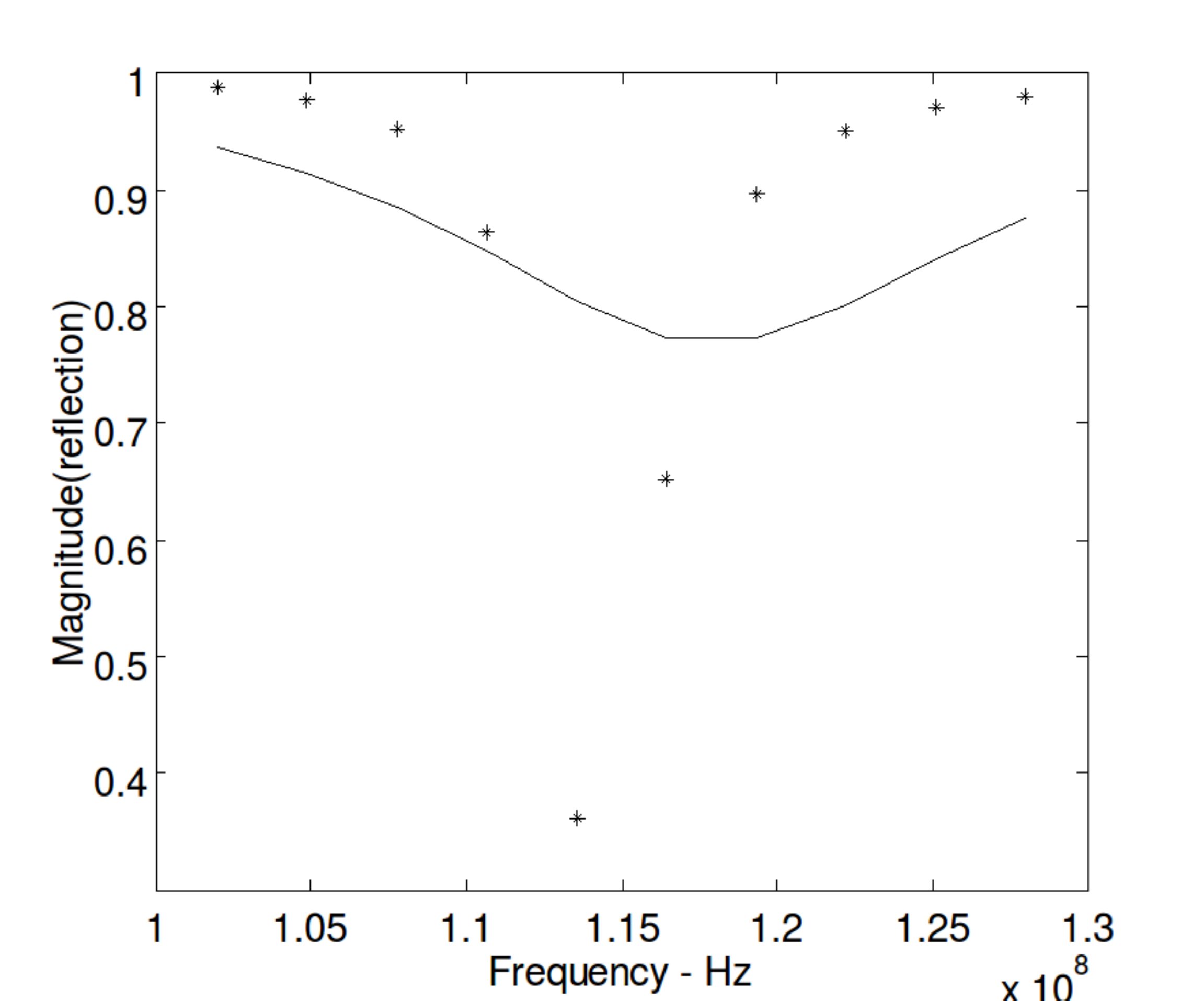}
\end{center}
  \caption{(Top) Error plots of simulation case for
    $s^{true}=10^{-3}$, $h^{true}=2$mm,  $R_{DGR}^{true}=10^{-2}$,
    $R_{WLR}^{true}=0.4$. Solid line is without open ended quarter
    resonator pair. (Bottom) Magnitude of reflection of capacitive 
    (+) and galvanic coupled (solid line) open ended coax resonators.} 
  \label{fig:error_4}
\end{figure}

\begin{figure}[ht]
\begin{center}
  \includegraphics[width=\linewidth]{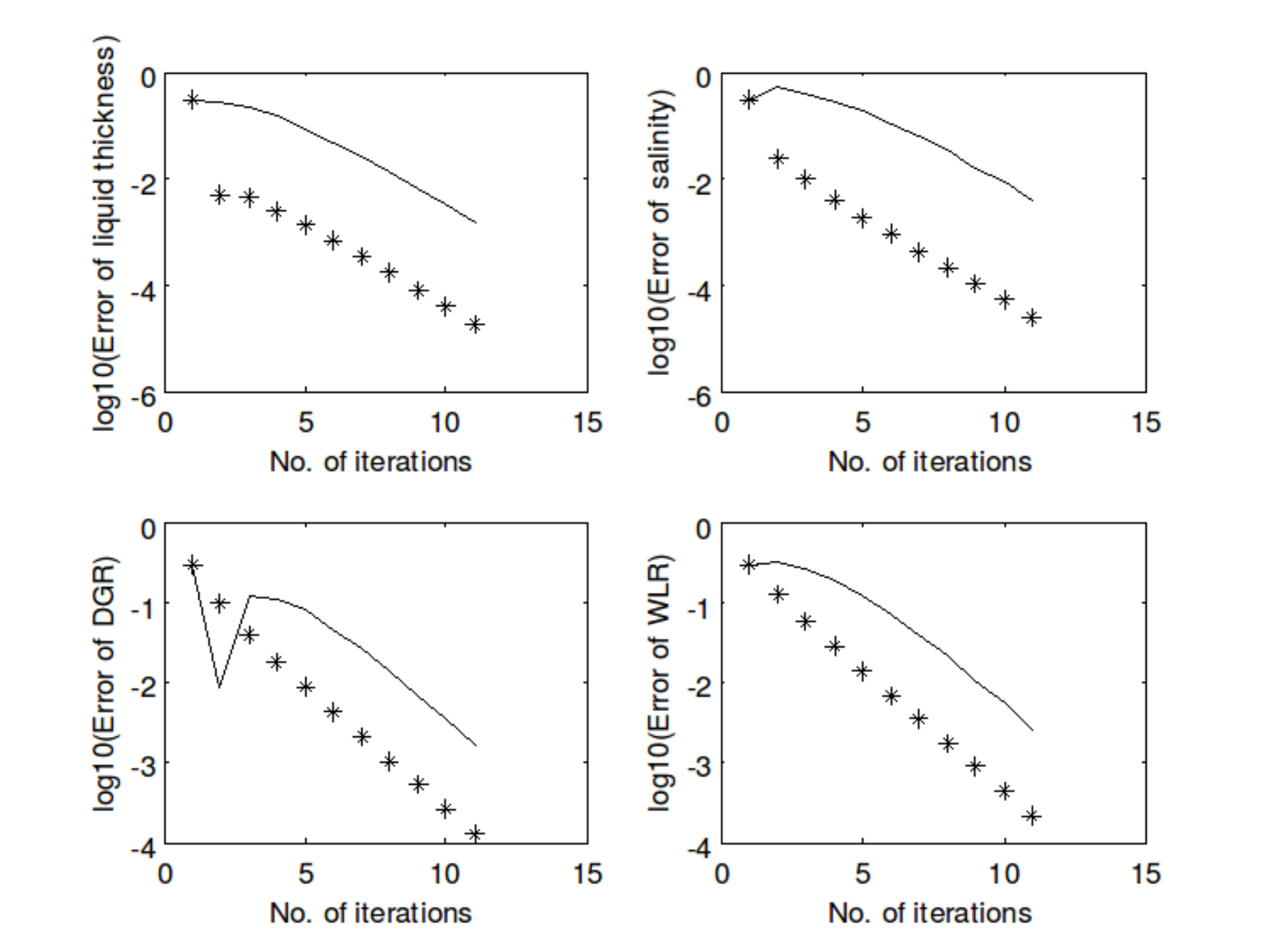}
  \includegraphics[width=0.7\linewidth]{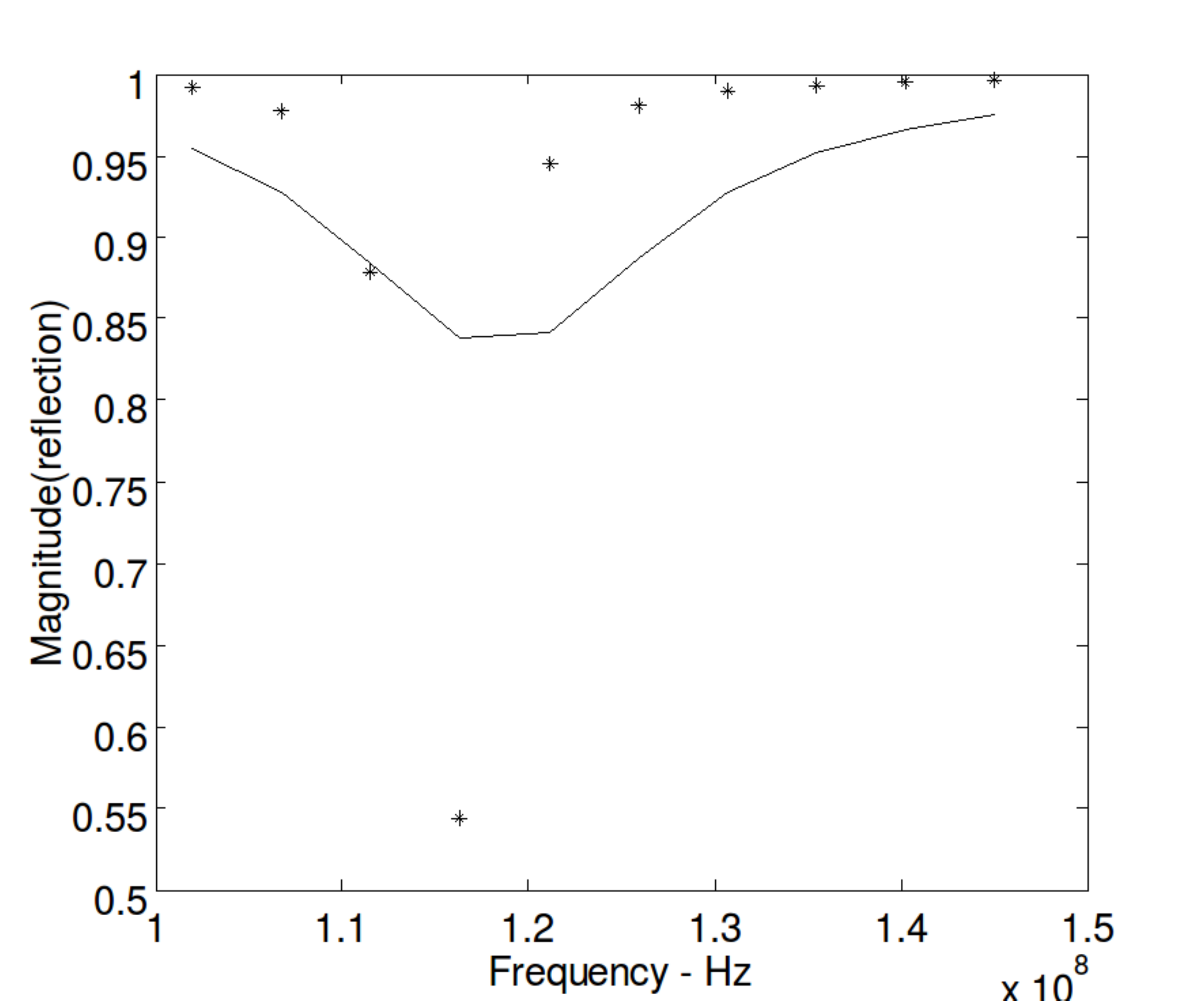}
\end{center}
  \caption{(Top) Error plots of simulation case for $s^{true}=0.1$,
    $h^{true}=2$mm, $R_{DGR}^{true}=10^{-2}$,
    $R_{WLR}^{true}=0.1$. Solid line is without open ended quarter
    resonator pair. (Bottom) Magnitude of reflection of capacitive (+)
    and galvanic coupled (solid line) open ended coax resonators.}
  \label{fig:error_5}
\end{figure}

Regarding the coupling of the open ended coax resonator pair, it is
noted that in the case of high salinity (typically
salinity $>\sim 3\%$ for $R_{WLR}>0.5$) one can obtain better salinity
sensitivity from amplitude changes with the galvanic coupled
resonator. This is demonstrated on Figures  \ref{fig:amp1} and \ref{fig:amp2} where we choose liquid thickness
$h=4$ mm, and WLR=1, amplitude change versus salinity (for
salinities from $4 \to 22\%$) is roughly a factor 2 larger for the
galvanic coupled resonator compared to the capacitive coupled
resonator. Thus, the
sensitivity of amplitude will be typically a factor 2 larger for the
galvanic coupled resonator except around the stagnation point at
salinity $\sim 16\%$. Note that the sensitivity is depending
somewhat on the coupling capacitance (in this work set to 10 pF).

\begin{figure}[ht]
\begin{center}
  \includegraphics[width=0.5\linewidth]{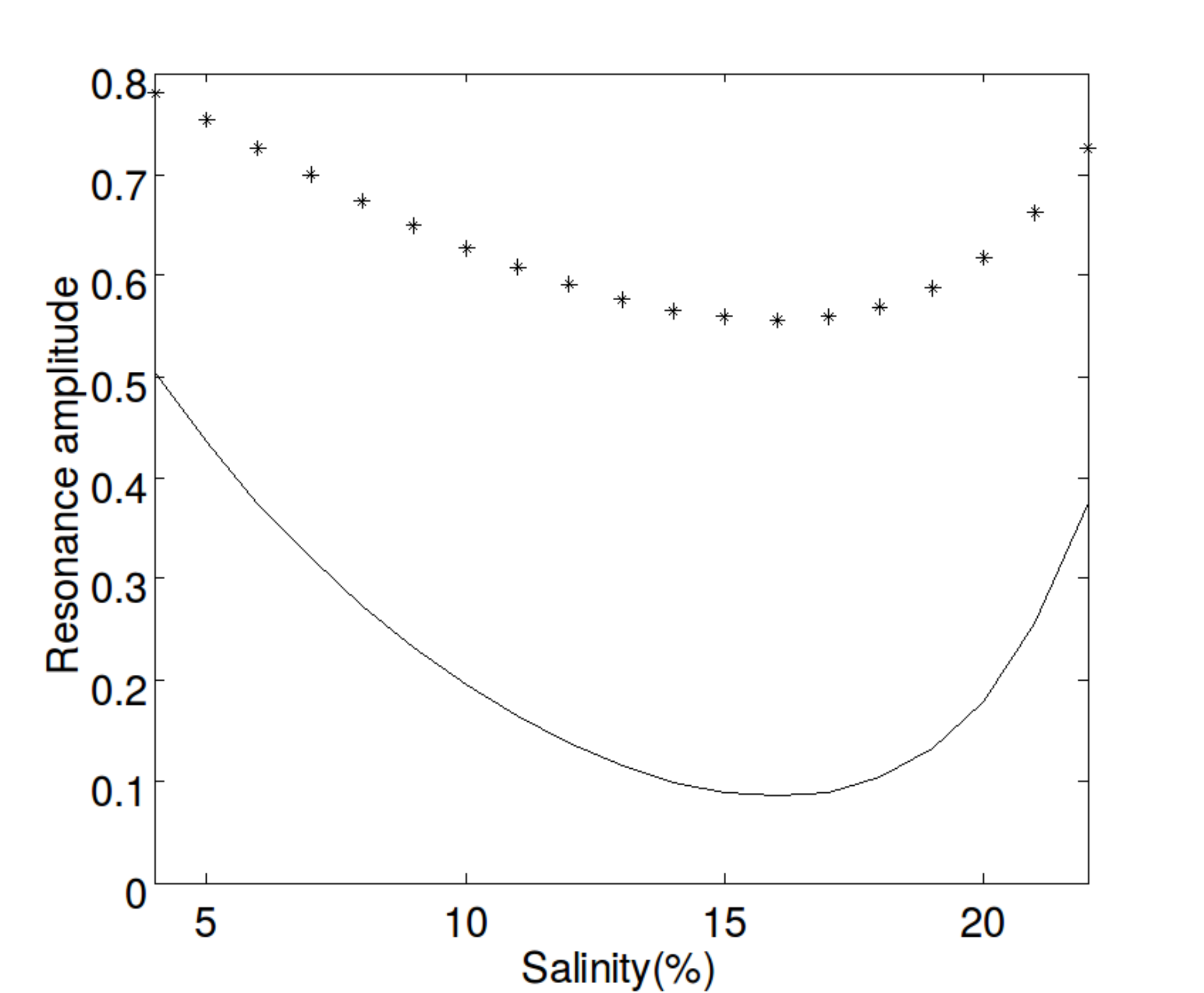}
\end{center}
  \caption{Comparison of amplitude for capacitive coupled (+) and
    galvanic coupled (solid line) open coax resonators at lowest
    resonance frequency versus salinity. Liquid thickness $h=4mm$,
    $R_{WLR}=1$ and $R_{DGR}=10^{-5}$.} 
  \label{fig:amp1}
\end{figure}

\begin{figure}[ht]
\begin{center}
  \includegraphics[width=0.5\linewidth]{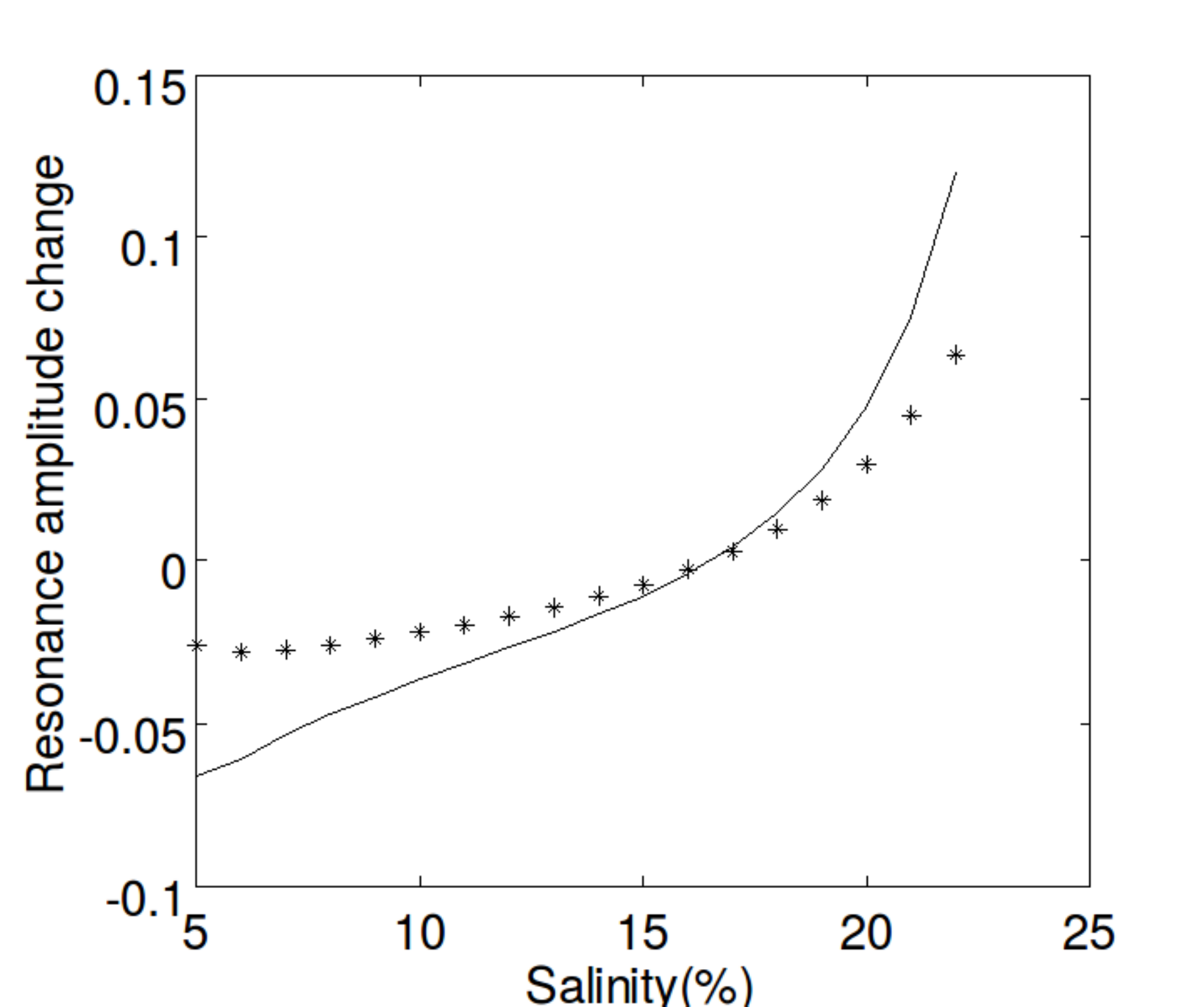}
\end{center}
  \caption{Comparison of amplitude change for capacitive coupled (+)
    and galvanic coupled (solid line) open coax resonators at lowest
    resonance frequency versus salinity. Liquid thickness $h=4$mm,
    $R_{WLR}=1$ and $R_{DGR}=10^{-5}$.}
  \label{fig:amp2}
\end{figure}

If we look at ``low'' salinity regime (salinity $0 \to \sim 3\%$ for
all WLR), the typical response of the capacitive coupled resonator is
the one of a nearly critical coupled resonator with sharply defined
skirts. The galvanic coupled resonator on the other hand, has a much
less pronounced resonance dip (see figure \ref{fig:reflection} for
simulation of reflection defined as reflected voltage/incident
voltage). A change in liquid thickness or change in WLR would lead
to a resonance frequency change and thus, the sensitivity of WLR or
liquid thickness change is greater for the capacitive coupled
resonator.

\begin{figure}[ht]
  \includegraphics[width=\linewidth]{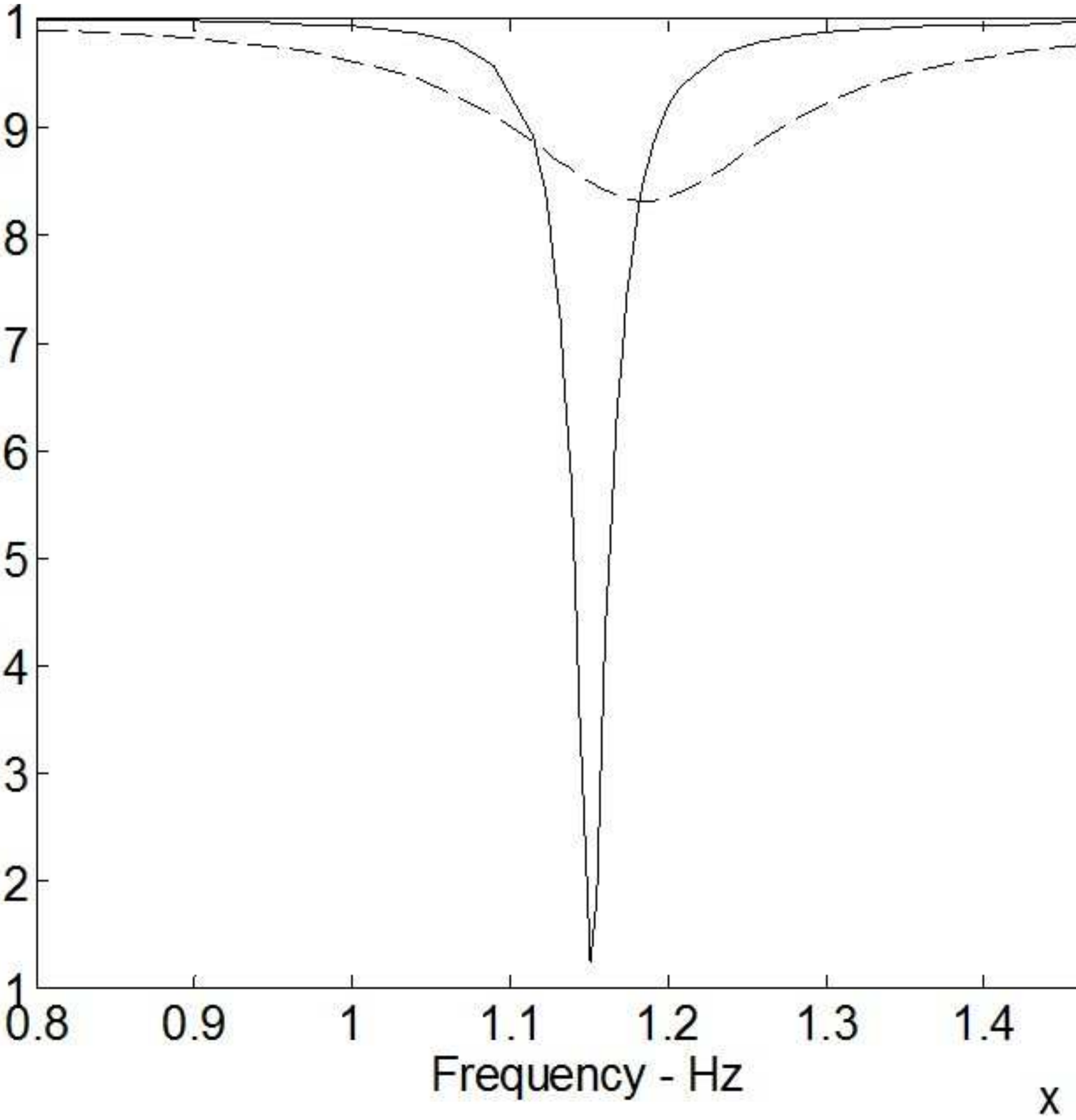}
  \caption{Simulation of reflection of capacitive coupled open coax
    resonator (solid line) and galvanic coupled open coax resonator
    (dashed line). $R_{WLR}=10^{-3}$, $R_{DGR}=10^{-2}$, $s=10^{-3}$
    and $h=0$.}  
  \label{fig:reflection}
\end{figure}

We denote the tunability as $\frac{f_0(h)-f_0(h + \delta h)}{f_0(h +
  \delta h)}$, where values of parameters are chosen as follows: $s= 10^{-7}$,
$R_{DGR}=0$, $R_{WLR} =1$.  In table \ref{tab:tunability1} is shown
the resonance frequency shift for a small liquid thickness change
$\delta h=0.5\mu m$. As seen in the table, the TE210 mode is shifted
twice as much as the TE110 modes. This is as expected, since
electrical field in TE210 mode is weaker in the center of the spool.

\begin{table}
  \centering
  \begin{tabular}{|p{0.3\linewidth}|p{0.2\linewidth}|p{0.2\linewidth}|}
    \hline
    Mode & Relative Tunability $h=1mm$ & Relative Tunability $h=2mm$ \\
    \hline
    TE110 & 3.701$\cdot 10^{-6}$ & 5.107$\cdot 10^{-6}$ \\
    \hline
    TE210 & 7.452$\cdot 10^{-6}$ & 1.293$\cdot 10^{-5}$ \\
    \hline
    Quarter wave & 1.520$\cdot 10^{-5}$ & 1.059$\cdot 10^{-5}$ \\
    \hline
  \end{tabular}
  \caption{The table shows the relative frequency change, by adding
    0.5$\mu m$ to the liquid film thickness $h$. Considering the $f_0$
    listed in table \ref{tab:tunability2}, these numbers represent
    frequency changes of the order $10^2 - 10^3$Hz.}
  \label{tab:tunability1}
\end{table}

As seen in table \ref{tab:tunability1}, the tunabilities for TE modes
and quarter wave coax are of the same order of magnitude, even though
quarter-wave resonator tunability decreases with increased liquid
layer $h$. It is the opposite for TE modes (at least in the
$h$-range shown here).

\begin{table}
  \centering
  \begin{tabular}{|p{0.325\linewidth}|p{0.25\linewidth}|p{0.25\linewidth}|}
    \hline
    Mode & Frequency, $f_0$ $h=1mm$ & Frequency, $f_0$ $h=2mm$  \\
    \hline
    TE110 & 1 041 453 582 & 1 032 491 153 \\
    \hline
    TE210 & 1 717 407 914 & 1 684 125 223 \\
    \hline
    Quarter wave, capacitive coupled & 112 188 018 & 110 530 685 \\
    \hline
  \end{tabular}
  \caption{The table shows the frequencies of the sensitivity analysed
    in table \ref{tab:tunability1}.}
  \label{tab:tunability2}
\end{table}
}

\section{\label{sec:discussion}Discussion}
We have presented a full wave transverse resonance model for a
circular cylindrical annular geometry. It was demonstrated
numerically that 4 unknown physical parameters could be extracted. If
we combine the transverse resonance model with the reflection data
from open ended quarter-wave resonators, we may improve convergence
and reducing error by a factor of 100.

It was also demonstrated, that the combination of a galvanic and
capacitive coupled open ended coax resonator renders higher
sensitivity for WLR and liquid thickness. This is
valid in low saline regime: salinity $< \sim 3\%$ for water continuous
liquid case or for any salinity where $R_{WLR} < 0.5$.

This frequency sensitivity improvement is due to capacitive coupled
open ended coax resonators sharper resonance pole skirts. For the high
saline regime (water-continuous and salinity $> \sim 3\%$), better sensitivity
(in amplitude change due to change in salinity) is obtained using the
galvanic coupled coax resonator.

\section*{Acknowledgements}
The authors are grateful for discussions with P.S. Kildal, Y. Yang,
Z. Sipus, P. Sl\"attman, H. Merkel and S. P. Hanserud.


\begin{thebibliography}{99}
 \raggedright
\bibitem{Eriksson}
  A. Eriksson, A. Deleniv and S. Gevorgian,
  Orientation and direct current field dependent dielectric properties of bulk single crystal SrTiO3 at microwave frequencies,
  \emph{J. Appl. Phys.}, 2003, 93, pp. 2848

\bibitem{Petersan}
  P.J. Petersan, S. M. Anlage,
  Measurement of Resonant Frequency and Quality Factor of Microwave Resonators: Comparison of Methods,
  \emph{J. Appl. Phys.},
  1998, 84, pp. 339


\bibitem{Eriksson_mode_chart}
  A. Eriksson, P.Linner and S. Gevorgian,
 Mode chart of electrically thin parallel-plate circular resonators,
  \emph{IEEE Proceedings-Microwaves Antennas and Propagation, issue: 1},
 2001, 148, pp. 51-55

\bibitem{Eriksson_res_tunell}
 A. Eriksson, A. Deleniv and S. Gevorgian,
  Resonant tunneling of microwave energy in thin film multilayer metal/dielectric structures, \emph{Microwave Symposium Digest, 2002 IEEE MTT-S International},
  2002, 3
 
\bibitem{Harrington}
 Roger F. Harrington,
  \emph{Time-Harmonic Electromagnetic Fields},
 Wiley-IEEE Press, 2001

\bibitem{Beilina}
 L. Beilina, A. Eriksson,
 Reconstruction of dielectric constants in a cylindrical waveguide,
  \emph{Springer Proceedings in Mathematics \& Statistics, Inverse Problems and Applications},
  2015, 120, pp. 97-109

\bibitem{Sipus}
  Z. {\v S}ipu{\v s}, P-S. Kildal, R.Leijon and M. Johansson,
  An Algorithm for Calculating Green's Functions of Planar, Circular Cylindrical, and Spherical Multilayer Substrates,
 \emph{ACES Journal}, 1998, 13 (3)

\bibitem{Gadani}
 D. H. Gadani et al.,
  Effect of Salinity on the dielectric properties of water,
  \emph{Indian Journal of Pure \& Applied Physics}, 2012, 50,
  pp. 405-410


\bibitem{Bruggeman}
  D. A. G. Bruggeman,
   Calculation of Various Physical Constants of Heterogeneous Substances,
  \emph{Ann. Phys.},
  1935, 32 (12)
 

\bibitem{mix_permitt_models}
  E. M. Kiley et al.,
  Applicability Study of Classical and Contemporary Models for Effective Complex Permittivity of Metal Powders,
  \emph{Journal of Microwave Power and Electromagnetic Energy},
  2012, 46, pp. 26-38

\bibitem{Baker_Jarvis}
  J. Baker-Jarvis et al.,
  Analysis of an Open-Ended Coaxial Probe with Lift-Off for Nondestructing Testing, \emph{IEEE Transactions on Instrument and Measurement}, 1994, 43 (5)
 

\bibitem{Klibanov_Bakushinsky_Beilina}
 M. V. Klibanov, A. B. Bakushinsky, L. Beilina,
  Why a minimizer of the Tikhonov functional is closer to the exact solution than the first guess, \emph{Journal of Inverse and Ill - Posed Problems}, 2011, 19,
  pp. 83-105


\bibitem{TihonovYagola}
 A. N. Tikhonov, A. V. Goncharsky, V. V. Stepanov and A. G. Yagola,
  \emph{Numerical Methods for the Solution of Ill-Posed Problems},
  London: Kluwer, 1995


\bibitem{BakushKokurinSmirnova}
  A. Bakushinsky, M. Kokurin, A. Smirnova,
  \emph{Iterative Methods for Ill-posed Problems},
 De Gruyter, 2011, Inverse and Ill-Posed Problems Series 54

\bibitem{Samar_Masterthesis}
  Samar Hosseinzadegan,
Iteratively regularized adaptive finite element method for reconstruction of coefficients in Maxwell's system,
  \emph{Master's thesis, Department of Mathematical Sciences, Chalmers, University of Gothenburg}, 2015 

\end{thebibliography}

\end{document}